\documentclass[aps,prl,twocolumn,superscriptaddress,showpacs,showkeys]{revtex4-1}
\usepackage{times}
\usepackage{color}
\usepackage{amssymb}
\usepackage{amsmath}
\usepackage{bm}
\usepackage{graphicx}
\usepackage{xfrac}
\usepackage{amsfonts}
\usepackage{hyperref}  
\hypersetup{breaklinks={true}}

\usepackage{xcolor,ulem}%

\renewcommand{\emph}{\textit}

\def\BigRoman{\uppercase\expandafter{\romannumeral\number\count 255 }}
\def\Romannumeral{\afterassignment\BigRoman\count255=}

\begin{document}

\title{Competition between group interactions and nonlinearity in voter dynamics on hypergraphs}
\author{Jihye Kim}
\affiliation{Department of Physics, Korea University, Seoul 02841, Korea}
\affiliation{School of Computational Sciences, Korea Institute for Advanced Study, Seoul 02455, Korea}
\author{Deok-Sun~Lee}
\email{deoksunlee@kias.re.kr}
\affiliation{School of Computational Sciences, Korea Institute for Advanced Study, Seoul 02455, Korea}
\affiliation{Center for AI and Natural Sciences, Korea Institute for Advanced Study, Seoul 02455, Korea}
\author{Byungjoon~Min}
\email{bmin@cbnu.ac.kr}
\affiliation{Department of Physics, Chungbuk National University, Cheongju, Chungbuk 28644, Korea}
\author{Mason~A.~Porter}
\email{mason@math.ucla.edu}
\affiliation{Department of Mathematics, University of California, Los Angeles, Los Angeles, CA 90095, USA}
\affiliation{Department of Sociology, University of California, Los Angeles, Los Angeles, CA 90095, USA}
\affiliation{Santa Fe Institute, Santa Fe, NM 87501, USA}
\author{Maxi~San~Miguel}
\email{maxi@ifisc.uib-csic.es}
\affiliation{Instituto de F\'isica Interdisciplinar y Sistemas Complejos, IFISC (CSIC-UIB), Campus Universitat Illes Balears, E-07122 Palma de Mallorca, Spain}
\author{K.-I.~Goh}
\email{kgoh@korea.ac.kr}
\affiliation{Department of Physics, Korea University, Seoul 02841, Korea}
\affiliation{Department of Mathematics, University of California, Los Angeles, Los Angeles, CA 90095, USA}


\begin{abstract}

Social dynamics are often driven by both pairwise (i.e., dyadic) relationships and higher-order (i.e., polyadic) group relationships,
which one can describe using hypergraphs. To gain insight into the impact of polyadic relationships on dynamical processes on networks,
we formulate and study a polyadic voter process, which we call the \textit{group-driven voter model} (GVM), that incorporates
the effect of group interactions by nonlinear interactions that are subject to a group (i.e., hyperedge) constraint.  
By examining the competition between nonlinearity and group sizes, we show that the GVM achieves consensus faster than standard voter-model dynamics, with an optimal minimizing exit time.
We substantiate this finding by using mean-field theory on annealed uniform hypergraphs with $N$ nodes, for which the exit time scales as ${\cal A}\ln N$, where the prefactor ${\cal A}$ depends both on the nonlinearity and on group-constraint factors. Our results reveal how competition between group interactions and nonlinearity shapes GVM dynamics. We thereby highlight the importance of such competing effects in complex systems with polyadic interactions.   
\end{abstract}


\date{\today}
\maketitle

{\it Introduction}---Individuals in society interact both in pairs and through various types of social groups (including families, clubs, and work colleagues) \cite{Olmsted1962,Wheelan1994}. 
Group (i.e., ``polyadic") interactions often are not merely structural units of a network; they also constitute functional units that drive dynamics through nonlinear effects \cite{Tyson1998,BP2019}.
Consequently, the traditional network framework --- which employs graphs and thus encodes group interactions 
as collections of pairwise (i.e., dyadic) interactions~\cite{bick2023} --- has a fundamental limitation.
To explicitly capture group interactions, one can employ ``higher-order'' (i.e., polyadic) network frameworks \cite{bick2023,Battiston2020,Bianconi_book,review2}.
There has been much recent work on dynamical processes on polyadic networks~\cite{review1,Iacopini2019,Skardal2019,st_onge-nonlinear-prl,St-Onge2022,deArruda2023,Carletti2023,Cencetti2023,nested,Burgio2024,HOC,Civilini2024} on a variety of systems, including opinion dynamics \cite{nonlinearcondition,majority,destroy,porter,adaptive-kuehn,adaptive}. 
However, researchers have not obtained
a generic understanding of the impact of group interactions on dynamical processes.

To gain insight into the impact of group interactions on opinion dynamics, we formulate and analyze a polyadic voter model. Voter models (VMs) \cite{voterbook} are both among the simplest models of social dynamics \cite{review_socio} and among the best-understood theoretical models of collective behavior of complex systems \cite{kinetic,dornic}. One can interpret the update rules of a VM~\footnote{There are several variants of ``the'' VM, depending on choices such as whether one selects nodes or edges at random, that have different qualitative dynamics~\cite{minireview}.} in terms of choosing between binary choices,  which we denote by
$\sigma = \mathbf{0}$ and $\sigma = \mathbf{1}$. We use the terms ``opinion" and ``state" interchangeably for the variable $\sigma$.
At each time step, a uniformly random node adopts the opinion of a uniformly random neighbor. (Henceforth, we use the term ``random" as a shorthand description for uniformly at random.) VMs have been studied for more than half of a century~\cite{holley1975}, and they
have been studied actively on traditional networks (i.e., graphs) for more than two decades~\cite{network2003,suchecki,sood2005,Vazquez2008PRL,sood2008,vaz,Masuda2014}.
VMs have also been extended in a variety of ways~\cite{minireview}.
However, few existing
studies account explicitly for group interactions \cite{adaptive-kuehn,adaptive}.

The framework of polyadic networks can help fill this gap by providing explicit structural models, such as hypergraphs and simplicial complexes \cite{bick2023}, to deal with group interactions. To 
incorporate group interactions into opinion dynamics, we use hypergraphs and generalize VM dynamics. To initiate our generalization, we first reformulate a traditional dyadic VM update rule by focusing on the role of edges. At each time step, a random node $i$ chooses one of its edges (i.e., links) at random, and it flips its state $\sigma_i$ to the state $\sigma_j$ of the adjacent node $j$ that is attached to the chosen edge if the node states are different.
In a dyadic network, each of these edges of a node $i$ is attached to exactly 1 other node.

\begin{figure}[t]
\centering
\includegraphics[width=.95\linewidth]{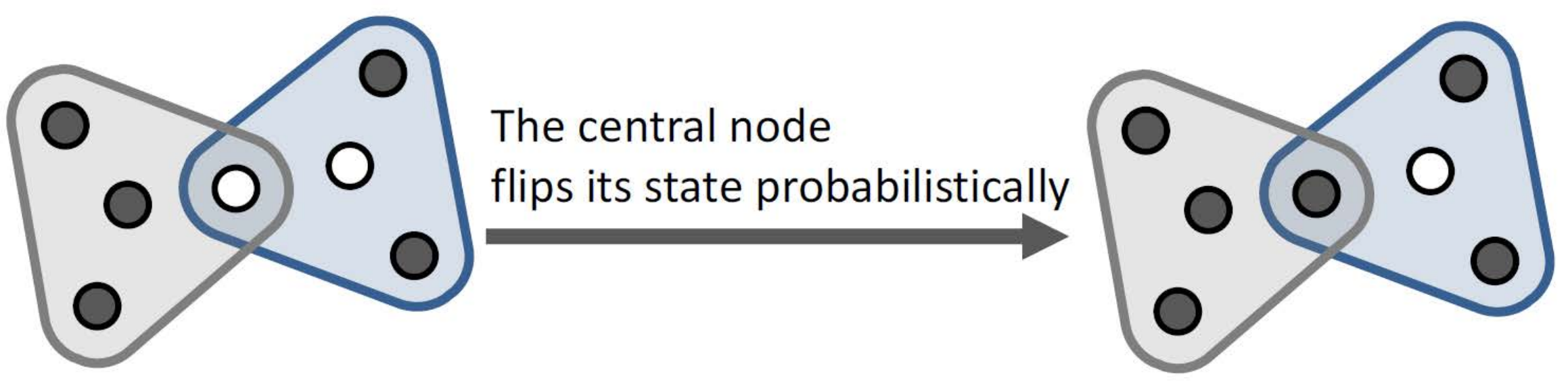}
\caption{Schematic illustration of our group-driven voter model (GVM) 
on a $4$-uniform hypergraph. The central node flips its state (that is, $\circ\to\bullet$) with a probability that depends on the GVM update rule. For example, for the simplicial GVM,
its flip probability is 1/2, as the central node has to select
the gray hyperedge to flip its state.
For the GVM with nonlinearity strength $q = 2$ and duplicate choices allowed, the central node's flip probability is instead $13/18$.
For the blue hyperedge, it has to select twice among its two black neighbors out of its three total neighbors.
}     
\label{fig:model}
\end{figure}


{\it Group-driven voter model}---In a hypergraph, a node $i$ can be adjacent to more than one other node via a hyperedge.
 Each node in a hyperedge with cardinality (i.e., ``size") $s$ is adjacent to $s - 1$ nodes.
This multiplicity leads to a broad spectrum of possibilities for dynamical processes on hypergraphs. To investigate these possibilities, we study a \textit{group-driven voter model} (GVM). 
At each time step, a random node $i$ considers adopting an opinion from one of its incident hyperedge $h$, which we choose randomly.
During the adoption process, node $i$ makes $q$ observations of states (i.e., opinions) of random nodes of hyperedge $h$.
One can either allow \cite{qvoter} or disallow \cite{centola-physa} duplicate choices of the same neighbor.
If the $q$ observed node states $\{\sigma_{j_1},\sigma_{j_2},\ldots,\sigma_{j_q}\,|\,j_p\in h\backslash \{i\}\}$ 
are unanimous and different from its own state
$\sigma_i$, then node $i$ flips its state (see Fig.~\ref{fig:model}) to match the observed state.

The GVM has two independent parameters: $q$ and $s$. The parameter $q$ accounts for nonlinear 
interactions \cite{Nowak1990}, which are absent in standard VMs
but have been considered in nonlinear variants of voter models \cite{vacillating,qvoter,confident,Kureh2020,Ramirez2024}. 
The parameter $s$ accounts for the effect of polyadic interactions. 
The GVM incorporates social reinforcement~\cite{power,centola-physa,kertesz-prl,synergy-juul}
via group interactions \cite{multibody_lam,st_onge-nonlinear-prl}, 
suggesting an explicit group-based origin of nonlinearity, which has been introduced in an ad hoc way in various dyadic variants of VMs~\cite{Spoof2019}, including a vacillating voter model \cite{vacillating}, a $q$-voter model~\cite{qvoter}, a confident voter model \cite{confident}, and a nonlinear voter model~\cite{Ramirez2024}. 

When $q = s - 1$ and duplicate choices are disallowed, the GVM captures the strongest group interactions, as it requires that all of the $s - 1$ nodes' states of a selected hyperedge are unanimous and different from node $i$'s state for node $i$ to flip its state. This requirement amounts to a ``simplicial rule", which was used in Refs.~\cite{Iacopini2019,ising2} to refer to polyadic
interactions that require unanimity of states. We thus refer to this variant as a ``simplicial GVM".
When $q = 1$ for all values of $s$, the GVM essentially reduces to a standard dyadic VM; it no longer experiences
the effects of polyadic interactions.
Additionally, the GVM on dyadic networks ($s = 2$) reduces to the standard VM for all values of $q$.
When $s = N$, {where $N$ is the number of nodes,} the GVM reduces to the noiseless $q$-voter model \cite{qvoter,Ramirez2024} on a fully-connected dyadic network. 
However, for networks that are not complete,
the correspondence is not exact due to the explicit group constraint.

\begin{figure}[t]
\centering
\includegraphics[width=.99\linewidth]{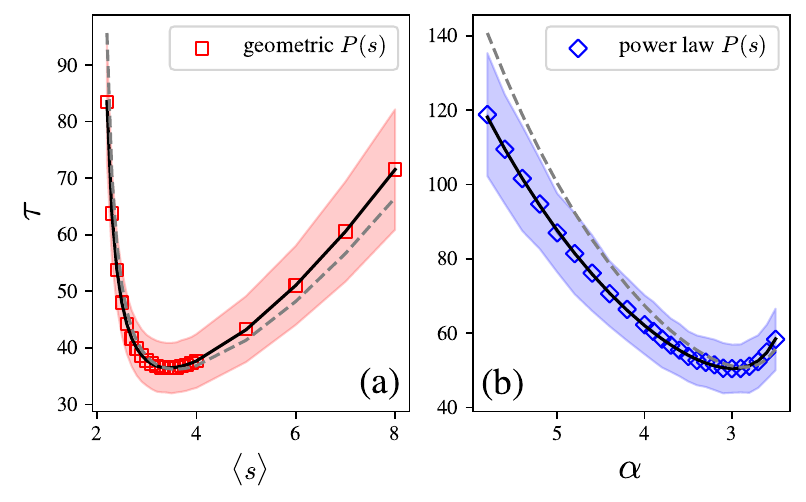}
\vskip -.3cm
\caption{The exit time $\tau$ for the simplicial GVM on annealed hypergraphs with hyperedge sizes that we obtain from (a) a geometric distribution with different mean hyperedge sizes $\langle s \rangle$  and (b) a power-law distribution with different power-law exponents $\alpha$. The symbols and shaded areas indicate the means and standard deviations, respectively, of $10^3$ independent Monte-Carlo simulations of the simplicial GVM with $N = 10^5$ nodes, and 
the curves indicate analytical results from a recursion relation (solid curves)
and a leading-order approximate solution (dotted curves). (See the Supplemental Material (SM) for the associated equations~\cite{SM}.) 
Both situations exhibit a notion of
optimality: With increasing $\langle s\rangle$ or decreasing $\alpha$, the exit time $\tau$ first decreases but eventually increases, achieving a minimum in the middle.}
\label{fig:tau_simplicial}
\end{figure}


To clearly observe the effect of groups, we consider the simplicial GVM. A basic property of voter dynamics is the exit time, which is the time that it takes to reach consensus (of either state). 
The exit time depends on the initial node states. Let
$\tau$ denote the exit time from a balanced initial condition, in which the same number of nodes are in each state.
In Fig.~\ref{fig:tau_simplicial}, we show the exit time $\tau$ for the simplicial GVM on ``annealed" hypergraphs, in which the nodes of a hyperedge are 
determined uniformly at random in each time step \cite{annealed}, 
with two different hyperedge-size distributions $P(s)$ --- a geometric distribution $P(s) = [(\langle s \rangle - 2)/(\langle s \rangle - 1)]^{s - 2}/(\langle s \rangle - 1)$ for $s \geq 2$
and a power-law distribution $P(s) = s^{-\alpha}/\sum_{\ell=2}^\infty \ell^{-\alpha}$ for $s\ge 2$ --- that are inspired by empirical 
data sets \cite{Patania2017,Benson2018,roh-jkps}. 
We compute the exit time $\tau$
as a function of the mean hyperedge size $\langle s \rangle$ (for the geometric distribution) and the power-law exponent $\alpha$ (for the power-law distribution) using 
Monte-Carlo (MC) simulations of the simplicial GVM on hypergraphs with $N = 10^5$ nodes \footnote{In the SM~\cite{SM}, we give the algorithmic details of our Monte-Carlo simulations.
Associated code is available at \url{https://github.com/JihyeKim2024/GVM.}}. 
In both cases, $\tau$ behaves nonmonotonically (see Fig.~\ref{fig:tau_simplicial}), so there are optimal values of $\tau$.
As groups of three or more nodes
begin to appear (i.e., $\langle s \rangle \gtrapprox 2$ or  
$3 \lessapprox \alpha < \infty$), consensus accelerates (i.e., $\tau$ becomes
 smaller). 
However, when group sizes are too large
(i.e., $\langle s\rangle \gg 2$ or $\alpha \lessapprox 3$), consensus decelerates. Therefore, there is an ``optimal" level of group interactions that leads to the fastest consensus (i.e., the smallest~$\tau$).

To gain theoretical insight into the origin of this optimality, we henceforth
 analyze the GVM with duplicate choices allowed on 
annealed $s$-uniform hypergraphs with $N$ nodes. 
We use
this setting because it allows us to derive
a series of concrete, informative analytical results.
For simplicity, we treat all nodes as equivalent; at each time step, we select nodes of a hyperedge uniformly at random with replacement.
In an $s$-uniform hypergraph, each hyperedge has the same size (i.e., the same number of nodes) $s$.
We consider the competition between our two independent parameters, $q$ and $s$, 
in the opinion dynamics. 
In this case, $q\geq s$ is also possible because we allow duplicate selections of the same neighboring node.


{\it Mean-field theory}---To theoretically understand the GVM dynamics, we use mean-field theory \cite{sood2008,Ramirez2024}.  
A key variable is the density $\rho(t)$, which is the fraction of nodes of a hypergraph in state $\mathbf{1}$ at time $t$.
In a time step, $\rho(t)$ can increase or decrease by $\delta\rho = 1/N$.
One can account for this change with the transition probability $R(\rho) \equiv P(\rho \rightarrow \rho + \delta\rho)$ that the number of nodes in state $\mathbf{1}$ increases by $1$ in a time step and the transition probability $L(\rho)\equiv P(\rho\rightarrow\rho - \delta\rho)$ that it
decreases by $1$ in a time step.
The probability of no change in $\rho$ in one time step is $1 - R(\rho) - L(\rho)$.
The rate equation for $\rho(t)$ is 
\begin{align}
	\dfrac{d\rho}{dt} = R(\rho) - L(\rho) \equiv v(\rho) \,,
\label{eq:rho}
\end{align}
where $v(\rho)$ is the drift function.

For an annealed $s$-uniform hypergraph, one can write~\cite{SM}
\begin{align}
	R(\rho) &= (1 - \rho)\sum_{n = 0}^{s - 1} {s - 1 \choose n}\rho^n (1 - \rho)^{s - 1 - n}\left(\frac{n}{s - 1}\right)^q \nonumber \\
		&= \left.\frac{(1 - \rho)}{(s - 1)^q}\left(\frac{d}{dr}\right)^q\left[(1 - \rho + \rho e^r)^{s - 1}\right]\right|_{r = 0} \,, \nonumber \\
	L(\rho) &= \rho\sum_{n = 0}^{s - 1} {s - 1 \choose n}\rho^n (1 - \rho)^{s - 1 - n}\left(1 - \frac{n}{s - 1}\right)^q \nonumber \\
		&= \left.\frac{\rho}{(s - 1)^q}\left(\frac{d}{dr}\right)^q\left[(\rho + e^r-\rho e^r)^{s - 1}\right]\right|_{r = 0} \,.
\label{eq:eqself}
\end{align}
In this mean-field approximation, the probability that a size-$s$ hyperedge has $n$ nodes in state $\mathbf{1}$ at time $t$ is 
${s - 1 \choose n}\rho^{n}(1 - \rho)^{s - 1 - n}$. 

The drift function $v(\rho)$ gives many useful insights about GVM dynamics. When $q = 1$ (i.e., for the standard VM), $v(\rho) = 0$ for all
$\rho$ because $R(\rho) = L(\rho) = \rho(1 - \rho)$~\cite{sood2008}. In this case, 
stochastic fluctuations enable finite-size systems to reach consensus.
For the generic GVM (i.e., when $q \geq 2$), the drift function $v(\rho)$ is no longer identically $0$.  
We show in the Supplemental Material (SM)~\cite{SM}
that Eq.~(\ref{eq:rho}) has three equilibrium points: $\rho = 0$, $\rho = 1$, and $\rho = 1/2$. The equilibrium points $\rho = 0$ and $\rho = 1$ are stable and correspond to 
consensus states with opinions $\mathbf{0}$ and $\mathbf{1}$, respectively.
Apart from finite-size fluctuations, the system eventually reaches the $\rho = 0$ consensus equilibrium whenever $\rho < 1/2$ because $v(\rho) < 0$. For $\rho > 1/2$, the system eventually reaches the consensus equilibrium $\rho = 1$. The unstable equilibrium point $\rho = 1/2$ has an equal mixture of the opinions $\mathbf{0}$ and $\mathbf{1}$. Drift towards a stable equilibrium point depends on the values of $q$ and $s$, which thereby play crucial roles in the GVM dynamics.
The drift function $v(\rho)$ of the GVM for $s = N$ reduces to that of the $q$-voter model on a fully-connected dyadic network~\cite{SM}.


{\it Sigmoidal exit probability}---Another key property of voter dynamics is the exit probability $\Phi(\rho)$, which is the probability to reach the 
opinion-$\mathbf{1}$ consensus state from the initial density $\rho$. 
From the preceding argument, we expect that the exit probability for  the generic GVM (i.e., for any $q \geq 2$) changes in a sigmoidal manner near
$\rho = 1/2$, with convergence to a step function in the thermodynamic limit $N \to \infty$, as has also been observed in numerical simulations of the $q$-voter model \cite{qvoter}.  
To confirm this expectation and 
 elucidate the group effect, we calculate 
$\Phi(\rho)$ explicitly for large but finite $N$. Following \cite{sood2008}, we set up the recursion relation
\begin{align}
	\Phi(\rho) &= R(\rho)\Phi(\rho + \delta\rho) + L(\rho)\Phi(\rho - \delta\rho) \nonumber \\
				&\quad \, + \left[1 - R(\rho) - L(\rho)\right]\Phi(\rho) 
\label{eq:eqp}
\end{align}
and Taylor-expand it in $\delta\rho = 1/N$ to second order to obtain a backward Kolmogorov equation
\begin{align}
	v(\rho)\dfrac{\partial{\Phi(\rho)}}{\partial{\rho}} + D(\rho)\dfrac{\partial^2{\Phi(\rho)}}{\partial{\rho^2}} = 0 \,,
\label{eq:Kolmogorov_P}
\end{align}
with a diffusion function $D(\rho) \equiv [R(\rho) + L(\rho)]/(2N)$ and boundary conditions $\Phi(0) = 0$ and $\Phi(1) = 1$. By symmetry, $\Phi(1 - \rho) = 1 - \Phi(\rho)$.


\begin{figure}[t]
\centering
\includegraphics[width=.9\linewidth]{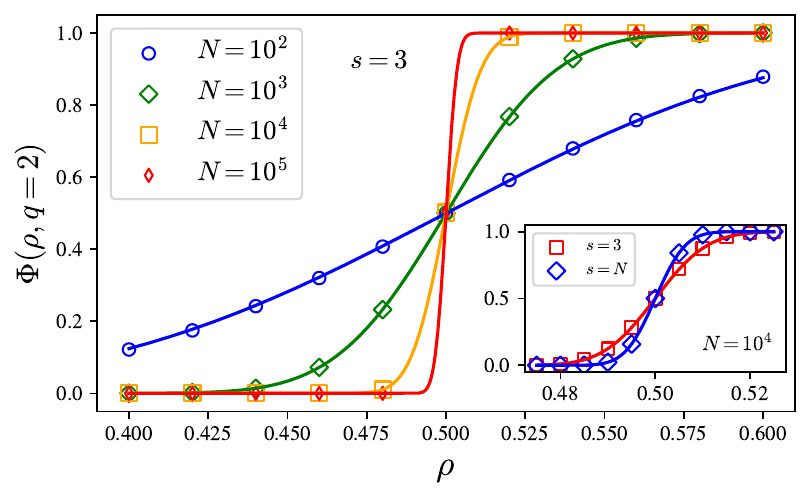}
\vskip -.3cm
\caption{The exit probability $\Phi(\rho)$ when the nonlinearity strength is $q = 2$ and the group size is $s = 3$. The solid curves are from the function in Eq.~(\ref{eq:exit_P_q2}), and the markers are means of $10^4$ independent MC simulations of the GVM on annealed $3$-uniform hypergraphs.
As $N$ increases, the sigmoid $\Phi$ converges to a step function. In the inset, we show $\Phi(\rho)$ for a fixed system size $N$ and different group sizes $s$. Convergence to a step function is slower for $s = 3$ than for $s = N$.
}
\label{fig:exitprob}
\end{figure}


To illustrate the effect of the nonlinearity, we compare the two simplest cases: $q = 1$ (i.e., the standard VM) and $q = 2$ (our GVM).
When $q = 1$, it is known that $\Phi(\rho) = \rho$ \cite{sood2008}, as one can view the dynamics as a diffusion process (i.e., $v(\rho) = 0$).
For $q = 2$, we solve Eq.~(\ref{eq:Kolmogorov_P}) explicitly to obtain~\cite{SM} 
\begin{align}
	\Phi(\rho) = \dfrac{1}{2} + \dfrac{\mathrm{erf}\left(\sqrt{\dfrac{2N(s - 2)}{s}}\left(\rho - \dfrac{1}{2}\right)\right)}{2\,\mathrm{erf}\left(\sqrt{\dfrac{N(s - 2)}{2s}}\right)} \,,
\label{eq:exit_P_q2}
\end{align}
where $\mathrm{erf}(\cdot)$ is the error function. 
In Fig.~\ref{fig:exitprob}, we plot $\Phi(\rho)$ when $q = 2$ and
$s = 3$. This expression
agrees with the results of our MC simulations.
This explicit closed-form confirmation demonstrates that the ``width'' $\Delta$ of the sigmoidal change across $\rho = 1/2$ scales as $\Delta \sim 1/\sqrt{N(s - 2)/s}$, illustrating both the finite-size effects (i.e., the dependence on $N$) and the group effect (i.e., the dependence on $s$). In particular, we see that convergence to a step function ``slows down'' for smaller group sizes $s$ when $q = 2$. See the inset of Fig.~\ref{fig:exitprob}.


{\it Logarithmic scaling of the exit time $\tau$ with hypergraph size $N$}---Let $T(\rho)$ denote the exit time for a general
initial density $\rho$ of nodes in state $\mathbf{1}$. Therefore, $\tau = T(\rho = 1/2)$.
Following a similar procedure as in our derivation of Eq.~\eqref{eq:Kolmogorov_P}
for the exit probability, 
we set up a recursion relation for the exit time $T(\rho)$. 
This yields the backward Kolmogorov equation~\cite{SM}
\begin{align}
	 v(\rho)\dfrac{\partial{T(\rho)}}{\partial{\rho}} + D(\rho)\dfrac{\partial^2{T(\rho)}}{\partial{\rho^2}} = -1 \,.
\label{eq:Kolmogorov_T}
\end{align}
For the standard VM (i.e., for $q = 1$), the drift term vanishes and we solve Eq.~(\ref{eq:Kolmogorov_T}) and obtain $\tau = N \ln 2$~
\cite{sood2008,SM}. However, for the GVM (i.e., for $q \geq 2$), it is typically not possible to solve Eq.~(\ref{eq:Kolmogorov_T}) analytically.
Nevertheless, one can numerically solve the recursion relation for $T(\rho)$ that is analogous to Eq.~(\ref{eq:eqp}). See
Eq.~(S21) in the SM \cite{SM}.

To proceed further analytically, 
we approximate Eq.~(\ref{eq:Kolmogorov_T}) by neglecting the diffusion term. We can do this because $D(\rho)/v(\rho)\sim 1/N$ as $N \rightarrow \infty$. We then integrate the resulting equation
to obtain the approximate exit time
\begin{align} \label{this}
	\tau \approx  \int_{\frac{1}{2}-\frac{1}{\sqrt{N}}}^{\frac{1}{N}} \frac{1}{v(\rho')}  d\rho' \,.
\end{align}
We have shifted the initial density by $1/\sqrt{N}$ from $1/2$ to exploit stochasticity and thereby avoid being trapped at the unstable equilibrium point. 
Under this approximation, we obtain to leading order in the hypergraph size $N$ that $\tau(N; s,q)\sim {\cal A}(s,q) \ln N$, where the prefactor ${\cal A}(s,q)$ 
depends on $s$ and $q$ for general $q \ge 2$ and $s \ge 3$. (See Eq.~(S33) in the SM~\cite{SM}.) 
One can attribute the logarithmic scaling of the exit time for the generic GVM 
to the fact
that $v(\rho) = 0$ has three simple roots in $[0,1]$.
The prefactor ${\cal A}(s,q)$ diverges for $s = 2$, as $\tau$ satisfies diffusive scaling $\tau \sim {\cal O}(N)$ for dyadic networks.

\begin{figure}[t]
\centering
\vskip -.05cm
\includegraphics[width=.96\linewidth]{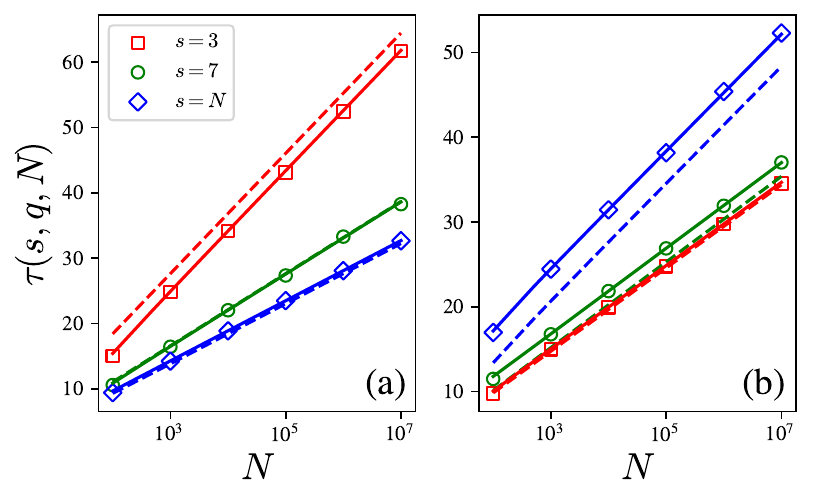}
\vskip -.3cm
\caption{Logarithmic scaling of the exit time $\tau(s,q,N)$ for the GVM on the hypergraph size $N$ for (a) nonlinearity strength $q = 2$ and (b) nonlinearity strength $q = 5$. The markers are means of $10^6$ (when $N \le 10^4$) or $10^3$ (when $N \ge 10^5$) independent MC simulations of the GVM
on annealed $s$-uniform hypergraphs. 
We obtain the solid curves from Eq.~(S21) in the SM~\cite{SM}, and we obtain the dashed curves from the leading-order solutions in Eqs.~(\ref{eq:q2_solution}, \ref{eq:q5_solution}).
The dashed and solid green curves ($s = 7$) in (a) almost
overlap.
} 
\label{fig:Ndependence_gala}
\end{figure}


It is insightful to show some explicit approximate expressions for
$\tau$ as examples. The leading-order expression of $\tau$ for $q = 2$ and $q = 5$ are~\cite{SM}
\begin{align}
	\tau(N,s;q = 2) &\sim \dfrac{2(s - 1)}{(s - 2)}\ln N \,,
\label{eq:q2_solution} \\
	\tau(N,s;q = 5)&\sim\dfrac{(s - 1)^4 \dfrac{(3s - 4)(s+1)}{\left(s^2 + 3s - 8\right)}}{s(s - 2)(s^2 - 2s + 2)}\ln N \,.
\label{eq:q5_solution} 
\end{align}
From Fig.~\ref{fig:Ndependence_gala}, we see
that the analytically-obtained logarithmic scaling of $\tau$ successfully explains the MC simulation results.
Figure~\ref{fig:Ndependence_gala} also reveals that 
the group effect can manifest distinctively for different nonlinearity strengths $q$. 
When $q = 2$, reaching consensus takes the longest time
for the smallest group size $s = 3$ [see Fig.~\ref{fig:Ndependence_gala}(a)]. By contrast, when $q = 5$, the longest consensus time occurs for the largest group size $s = N$ [see Fig.~\ref{fig:Ndependence_gala}(b)]. 


\begin{figure}[t]
\centering
\vskip -.1cm
\includegraphics[width=.97\linewidth]{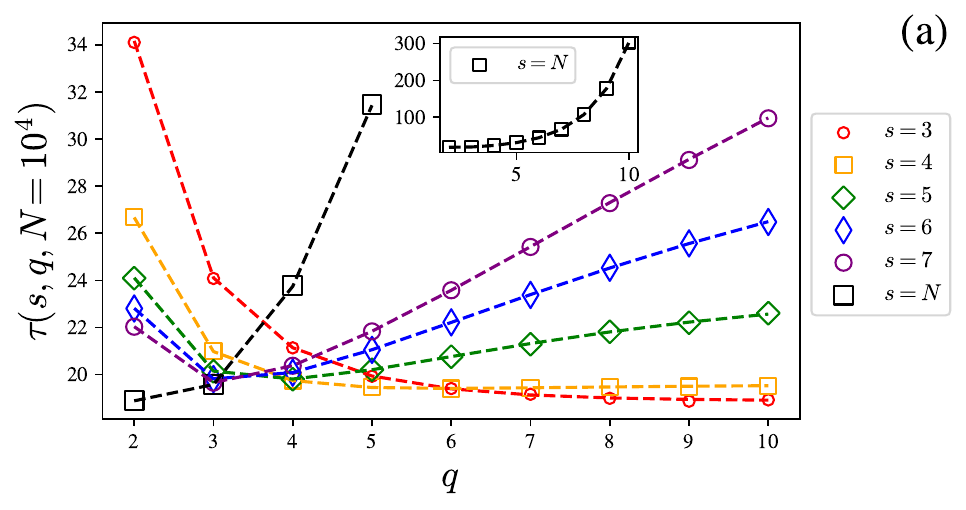}
\vskip -.15cm
\includegraphics[width=.99\linewidth]{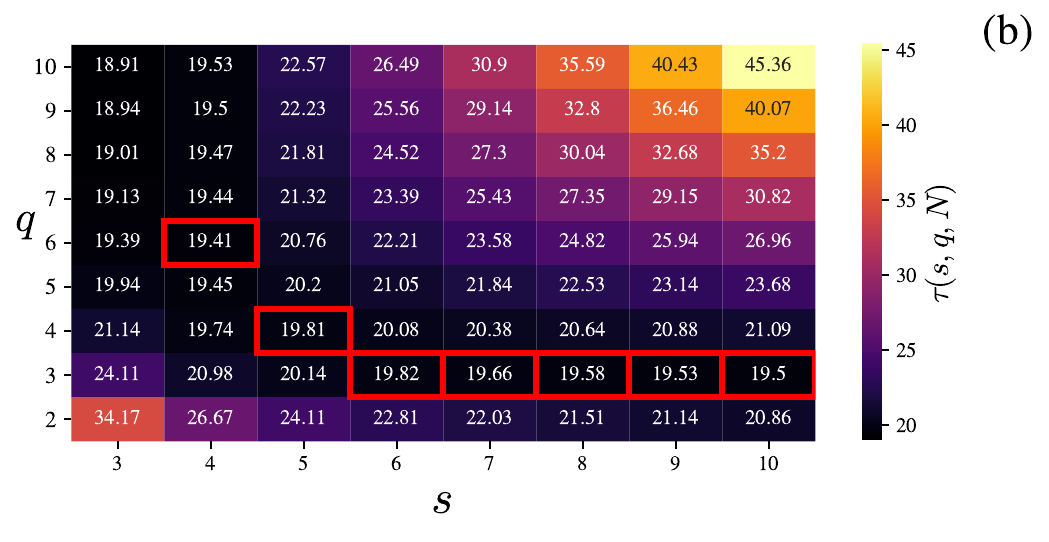}
\vskip -.3cm
\caption{
(a) The dependence on the nonlinearity strength $q$ of the exit time $\tau$ for the GVM for different values of the group size $s$ for hypergraphs with $N = 10^4$ nodes. The markers are means of $10^4$ independent MC simulations of the GVM on annealed $s$-uniform hypergraphs, and the analytical curves are from numerical solutions of the recursion relation (S21) in the SM~\cite{SM}.
When $s = N$, the exit time $\tau$ grows exponentially quickly with $q$, making it problematic to depict it along with other cases. In the inset, we show the curve
for $s = N$ in an extended $q$ range. 
(b) The heat map for $\tau(s,q,N)$ that we obtain from numerical solutions of the recursion relation (S21).
For a given $s$, the cell with the red border has the optimal $\tau$.
}
\label{fig:qdependence_gala}
\end{figure}

{\it Optimality in the exit time $\tau$}---To further examine the 
interplay between the nonlinearity and group effects on the exit time $\tau$, 
we investigate how varying the nonlinearity strength $q$ affects the GVM dynamics for specified values of the group size $s$ and hypergraph size $N$.
In the absence of the group constraint (i.e., $s = N$), the leading-order expression of the exit time $\tau$ is 
\begin{align} 
	\tau(q;N,s = N) \sim  \left(1 + \frac{2^{q - 2}}{q - 1}\right)\ln N \,.
\label{eq:tau_s=N}
\end{align} 
This expression also applies to the
$q$-voter model on a complete dyadic graph.
Because $\tau$ increases with $q$, a stronger nonlinearity decelerates
consensus. [See Fig.~\ref{fig:qdependence_gala}(a) and its inset.]
By contrast, with the most-constraining groups (i.e., $s = 3$),
the exit time $\tau$ decreases monotonically with $q$. To leading order, $\tau(q;N,s = 3) \sim \frac{2^q}{2^{q - 1} - 1}\ln N$. A stonger nonlinearity accelerates consensus. [See the red curve in Fig.~\ref{fig:qdependence_gala}(a).] 

There is a nontrivial tradeoff between these extreme situations.
As we can see in Fig.~\ref{fig:qdependence_gala}(a), for a given hyperedge size $s$, there is an optimal nonlinearity strength $q^*$ with the minimum exit time $\tau$.
We systematically investigate the tradeoff 
for many values of $s$ and $q$ [see Fig.~\ref{fig:qdependence_gala}(b)].  These computations reveal the global landscape and optimality of GVM dynamics.

We now explain why we observe optimality.
From the inset of Fig.~\ref{fig:qdependence_gala}(a), we see that considering the opinions of exactly $2$ neighbors
is the most efficient way to achieve consensus. 
Increasing $q$ in Eq.~(\ref{eq:tau_s=N}) with $s = N$ reduces the probability that neighbors have unanimous opinions, 
which in turn decreases 
the drift function $v(\rho)$ and decelerates the approach to consensus.
Because $q$ grows logarithmically with $N$ [i.e., $q\sim{\cal O}(\ln N)$], the logarithmic scaling of $\tau$ in Eq.~\eqref{eq:tau_s=N} 
eventually becomes a linear scaling $\tau\sim{\cal O}(N)$, which is comparable to the diffusive scaling
for $q = 1$. 
However, the probability that a node consults the same neighbor twice (instead of consulting 2 different neighbors) for $q = 2$ is $1/(s - 1)$, which is not negligible for small $s$.
In this situation, a node only consults the opinion of
$1$ neighbor, so it again effectively follows diffusive dynamics. 
More generally, the probability of diffusive dynamics from consulting just $1$ neighbor increases with decreasing $q$.
Therefore, there is a ``sweet spot" $q^*$ that minimizes $\tau$ between the two diffusive-dynamics maxima. That is,
$2 <  q^* < {\cal O}(\ln N)$.
The case $s = 3$ is a notable exception.
 When $s = 3$, the
maximum number of different neighbors is $2$, 
so $\tau$ decreases indefinitely (although slowly) as $q$ increases.

When $q$ is fixed, increasing $s$ towards $N$ reduces the probability of unanimity, as the number of distinct neighbors that are chosen increases, and decreasing $s$ towards $2$ increases the probability of consulting just $1$ neighbor. Both  situations lead to an increase in the exit time $\tau$. Therefore, there exists an optimal $\tau$.
An equivalent explanation of
the presence of optimality in the simplicial GVM (with $q = s - 1$ and duplicate choices disallowed)
in Fig.~\ref{fig:tau_simplicial} is as a
competition between 
diffusive dynamics from 
dyadic edges (which dominates as $\langle s \rangle \downarrow 2$ and $\alpha \to \infty$) 
and the small probability of unanimity in large hyperedges (which dominates as $\langle s \rangle \to \infty$ and $\alpha \downarrow 2$). The exit time $\tau$ increases as one approaches either of these limits, so there is an optimal $\tau$.


{\it Conclusions}---We formulated and analyzed a group-driven voter model (GVM) that accounts for the effects of both polyadic interactions and nonlinear interactions within groups.  A larger nonlinearity strength $q$ leads 
to faster consensus in the GVM than in conventional VMs, which exhibit diffusive dynamics.
This acceleration of consensus formation depends on
the interplay between the nonlinearity strength $q$ and the group size $s$ of hypergraphs. Through mean-field calculations and Monte-Carlo simulations, we demonstrated that the exit time scales logarithmically with system size and that there is an optimal value $q^*$ of the nonlinearity strength $q$ that minimizes the exit time.
This optimality emerges from a competition between diffusive dynamics when both $q$ and $s$ are small and a slow drift when both $q$ and $s$ are large.  This emergent group effect cannot arise in dyadic networks.

We also apply our analytical approach to several variants of our GVM (see the SM \cite{SM}): a
simplicial GVM, a GVM without allowing duplicate choices, and a GVM with edge-update dynamics in which we simultaneously update the opinions of all nodes that are attached to a hyperedge. In all of these cases, the exit time scales logarithmically with system size, illustrating the robustness of our main theoretical results~\cite{SM}. To further examine the robustness of our results, it is also important to consider additional phenomena.
For example, our analysis did not account for heterogeneities in the degree distribution (where the degree of a node is the number of hyperedges it is in).
Our preliminary Monte-Carlo calculations \cite{SM} illustrate that the degree distribution can influence
 the exit time. Therefore, it will be useful to generalize our mean-field framework to a degree-based mean-field theory~\cite{sood2005,sood2008,vaz} to study their effects {analytically}.
It will also be useful to extend our GVM to encompass more realistic aspects (see, e.g., \cite{minireview}) of opinion dynamics.
There have been studies of optimal group and team sizes in 
social psychology \cite{groupsize1,groupsize2}, and further studies of our GVM and its generalizations may yield interesting insights about these phenomena.

\begin{acknowledgments}

{\it Acknowledgments}---We thank Sid Redner for useful conversations and suggestions.
This work was supported in part by National Research Foundation of Korea (NRF) grants funded by the Korea government (MSIT) [No.~2020R1A2C2003669 (K-IG) and No.~2020R1I1A3068803 (BM)], by a KIAS Individual Grant [No.~CG079902] at Korea Institute for Advanced Study (D-SL), 
by the National Science Foundation (grant number 
1922952) through their program on Algorithms for Threat Detection (MAP), 
and by Agencia Estatal de Investigaci\'on (AEI, MCI, Spain) MCIN/AEI/10.13039/501100011033 and Fondo Europeo de Desarrollo Regional (FEDER, UE) under Projects APASOS [PID2021-122256NB-C21] and the Maria de Maeztu Program for Units of Excellence in R\&D grant [CEX2021-001164-M] (MSM).

\end{acknowledgments}


\nocite{*}


\bibliography{GVM-R1-reflist}


\end{document}


\title{Supplemental Material for ``Competition between group interactions and nonlinearity in voter dynamics on hypergraphs''}
\author{Jihye Kim}
\affiliation{Department of Physics, Korea University, Seoul 02841, Korea}
\affiliation{School of Computational Sciences, Korea Institute for Advanced Study, Seoul 02455, Korea}
\author{Deok-Sun Lee}
\email{deoksunlee@kias.re.kr}
\affiliation{School of Computational Sciences, Korea Institute for Advanced Study, Seoul 02455, Korea}
\affiliation{Center for AI and Natural Sciences, Korea Institute for Advanced Study, Seoul 02455, Korea}
\author{Byungjoon Min}
\email{bmin@cbnu.ac.kr}
\affiliation{Department of Physics, Chungbuk National University, Cheongju, Chungbuk 28644, Korea}
\author{Mason A. Porter}
\email{mason@math.ucla.edu}
\affiliation{Department of Mathematics, University of California, Los Angeles, Los Angeles, CA 90095, USA}
\affiliation{Department of Sociology, University of California, Los Angeles, Los Angeles, CA 90095, USA}
\affiliation{Santa Fe Institute, Santa Fe, NM 87501, USA}
\author{Maxi San Miguel}
\email{maxi@ifisc.uib-csic.es}
\affiliation{Instituto de F\'isica Interdisciplinar y Sistemas Complejos, IFISC (CSIC-UIB), Campus Universitat Illes Balears, E-07122 Palma de Mallorca, Spain}
\author{K.-I.~Goh}
\email{kgoh@korea.ac.kr}
\affiliation{Department of Physics, Korea University, Seoul 02841, Korea}
\affiliation{Department of Mathematics, University of California, Los Angeles, Los Angeles, CA 90095, USA}

\maketitle


This supplemental material for ``Competition between group interactions and nonlinearity in voter dynamics on hypergraphs'' consists of three major sections.


\tableofcontents
\makeatletter
\let\toc@pre\relax
\let\toc@post\relax
\makeatother 


\section{Monte-Carlo simulations of a GVM}

In this section, we provide algorithmic details of our Monte-Carlo (MC) simulations. Associated code is available at \url{https://github.com/JihyeKim2024/GVM}.


\subsection{Simplicial GVM on annealed hypergraphs}

In Fig.~2 of the main manuscript, we showed results of MC simulations of the simplicial GVM on annealed hypergraphs with $N$ nodes and hyperedge-size distribution $P(s)$. In an annealed hypergraph, the elements of a hyperedge are not fixed (i.e., ``quenched"); instead, one determines them randomly at each time step.
Additionally, we assume that all nodes are equivalent.
Each MC step has the following three stages: 
\begin{itemize}
\item[(i)] We select a node $v$ uniformly at random with probability $1/N$. 
\item[(ii)] We draw a random number $s$ from the probability distribution $sP(s)/\langle s \rangle$, where $\langle s \rangle = \sum_{s = 2}^{\infty} sP(s)$ is the mean hyperedge size.
The probability that a randomly selected hyperedge of node $v$ has size $s$
is proportional to $sP(s)$.
We select $s - 1$ distinct nodes uniformly at random from the $N - 1$ other nodes (i.e., excluding $v$ itself) of the hypergraph to form a hyperedge $h$.
\item[(iii)] The node $v$ flips its state $\sigma_v$ if and only if the states of all $s - 1$ other nodes in the selected hyperedge $h$ are unanimous and different from $\sigma_v$.
\end{itemize}


\subsection{GVM on annealed $s$-uniform hypergraphs}\label{stages}

In Figs.~3--5 of the main manuscript, we showed results of MC simulations of the GVM on annealed $s$-uniform  
hypergraphs with $N$ nodes. An $s$-uniform hypergraph is a hypergraph in which every hyperedge has the same cardinality (i.e., size) $s$. 
We again assume that all nodes are equivalent.
Each MC step has the following three stages: 
\begin{itemize}
\item[(i)] We select a node $v$ uniformly at random with probability $1/N$.
\item[(ii)] We select $s - 1$ distinct nodes uniformly at random from the other $N - 1$ nodes (i.e., excluding $v$ itself) of the hypergraph to form
a hyperedge $h$. 
\item[(iii)] We select a node other than $v$ from the hyperedge $h$ uniformly at random from the $s - 1$ remaining nodes, and we record the state $\sigma$ of this node.
 We repeat this process $q - 1$ times for a total of $q$ independent instances of this process.  
The node $v$ flips its state $\sigma_v$ if and only if the $q$ states 
are unanimous and different from $\sigma_v$. 
\end{itemize} 
In stage (iii), one can either allow or disallow
duplicate selections of the same neighboring node. The GVM in the main manuscript does allow duplicate selections. 
In Sec.~\ref{sec-s3b}, we consider a variant GVM in which we do not allow duplicate selections.


\section{Detailed derivations of our main analytical results} 

In this section, we give detailed derivations of our main analytical results for the GVM in the main manuscript. In this GVM, a node consults $q$ neighboring opinions with duplicate selections allowed. 


\subsection{Transition probabilities for general $P(s)$}
\label{transition}

\begin{figure}
\includegraphics[width=0.45\linewidth]{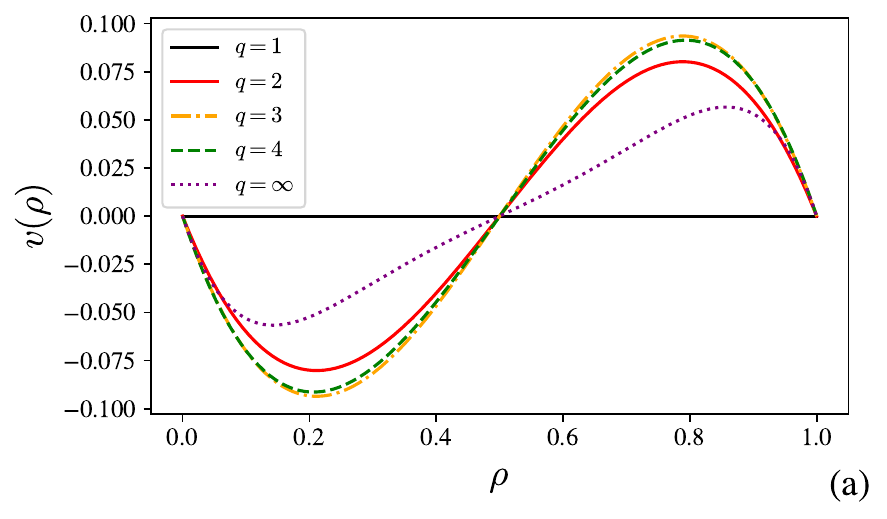}
\includegraphics[width=0.45\linewidth]{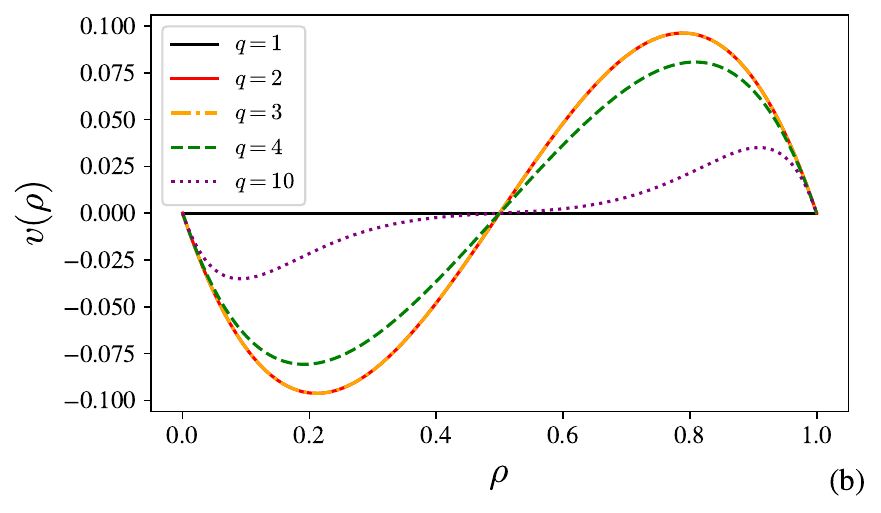}
\caption{The drift function $v(\rho) = R(\rho) - L(\rho)$ for different values of $q$ when (a) $s = 7$ and (b) $s = N \rightarrow \infty$. The curves for $q = 2$ and $q = 3$ in (b) completely overlap. 
}
\label{eq:v_s=N}
\end{figure}

To track the time evolution of the fraction $\rho(t)$ of nodes in
state $\mathbf{1}$ at time $t$ in a hypergraph, we consider the transition probabilities $R(\rho) \equiv P (\rho \to \rho + 1/N)$ (i.e., the ``raising operator") and $L(\rho) \equiv P(\rho\to\rho - 1/N)$ (i.e., the ``lowering operator")~\cite{sood-redner}.
The probability that a selected hyperedge in stage (ii) has size $s$ is proportional to $sP(s)$, 
so the mean-field expression for the raising operator $R(\rho)$ is
\begin{align}
	R(\rho) &= \dfrac{(1-\rho)}{\sum_s sP(s)}\sum_s sP(s)\sum_{n=1}^{s - 1} \dfrac{(s-1)!}{n!(s - 1 - n)!}\rho^n(1 - \rho)^{s - 1 - n}\left(\frac{n}{s - 1}\right)^q
\nonumber \\
	&= \dfrac{\rho(1-\rho)}{\sum_s sP(s)}\sum_{s} sP(s)R_s(\rho)\,,
\label{eq:R}
\end{align}
where 
\begin{align}
	R_s(\rho)=\sum_{n=1}^{s-1} \dfrac{(s-2)!}{(n-1)!(s-1-n)!}\rho^{n-1} (1-\rho)^{s-1-n}\left(\dfrac{n}{s-1}\right)^{q-1}\,. 
\label{eq:Rs}
\end{align}
For an annealed $s$-uniform hypergraph, Eq.~(\ref{eq:R}) reduces to $R(\rho)$ of Eq.~(2) of the main manuscript.
We express
Eq.~(\ref{eq:Rs}) in terms of
$\rho$, $s$, and $q$ using the relation
\begin{align}
	n^{q-1} = 1 + \sum\limits_{r = 1}^{q - 1} A_{r,q} \prod_{l = 1}^r (n - l)\,,
\label{eq:n_q}
\end{align}
with positive integers 
\begin{equation}
	A_{r,q} \equiv \left[\dfrac{(r+1)^{q-1}-r^{q}+r-1}{r!}+\mathbf{1}_{r\geq 3}\sum\limits_{l=2}^{r-1}\dfrac{(r-l+1)^{q-1}-1}{l!(r-l)!}(-1)^l\right]\,, 
\end{equation}	
where the indicator symbol $\mathbf{1}_{r\geq 3}$ has the value $1$ when $r\geq 3$ and has the value $0$ otherwise. Note that $A_{r,q} = 0$ for $r\geq q$ and that $A_{q - 1,q} = 1$.
Inserting Eq.~(\ref{eq:n_q}) into Eq.~(\ref{eq:Rs}) yields
\begin{align}
	R_s(\rho)&=\dfrac{1}{(s-1)^{q-1}}\left[1+ \sum_{n=1}^{s-1} \sum_{r=1}^{q-1} \dfrac{(s-2)!}{(n-1)!(s-1-n)!}\rho^{n-1} (1-\rho)^{s-1-n}A_{r,q}\prod_{l=1}^r (n-l)\right]
\nonumber \\
&=\dfrac{1}{(s-1)^{q-1}}+\dfrac{1}{(s-1)^{q-1}}\sum_{n=2}^{s-1} A_{1,q}\dfrac{\rho(s-2)(s-3)!}{(n-2)!(s-1-n)!}\rho^{n-2} (1-\rho)^{s-1-n}
\nonumber \\
&\quad+\dfrac{1}{(s-1)^{q-1}}\sum_{n=3}^{s-1} A_{2,q}\dfrac{\rho^{2}(s-2)(s-3)(s-4)!}{(n-3)!(s-1-n)!}\rho^{n-3} (1-\rho)^{s-1-n} + \cdots
 \nonumber \\ 
&\quad+\dfrac{1}{(s-1)^{q-1}}\sum_{n=q}^{s-1} A_{q-1,q}\dfrac{\rho^{q-1} (s-q-1)! \prod_{l=1}^{q-1} (s-1-l)}{(n-q)!(s-1-n)!}\rho^{n-q} (1-\rho)^{s-1-n}
\nonumber \\
&=\dfrac{1}{(s-1)^{q-1}}\left[1+A_{1,q}\rho(s-2)+A_{2,q}\rho^2 (s-2)(s-3) + \cdots + A_{q-1,q}\rho^{q-1}\prod_{l=1}^{q-1} (s-1-l)\right] 
\nonumber \\
	&=\dfrac{1}{(s-1)^{q-1}}\left[1+\sum_{r=1}^{q-1} A_{r,q}\rho^{r}\prod_{l=1}^{r}(s-1-l)\right]\,.
\end{align}
Therefore, 
\begin{align}
	R(\rho)&=\dfrac{\rho(1-\rho)}{\sum_s sP(s)}\sum_s \dfrac{sP(s)}{(s-1)^{q-1}} \left[1+\sum_{r=1}^{q-1} A_{r,q}\rho^{r}\prod_{l=1}^{r}(s-1-l)\right]\,.
\label{eq:reduce_R}
\end{align}
When $q \geq s$, the leading term in the square brackets of Eq.~\eqref{eq:reduce_R}
is $A_{s-2,q}\rho^{s-2}(s-2)!$.
The lowering operator $L(\rho)$  satisfies $L(\rho) = R(1-\rho)$, so
\begin{align}
	L(\rho) &= \dfrac{\rho(1-\rho)}{\sum_s sP(s)}\sum_s \dfrac{sP(s)}{(s-1)^{q-1}} \left[1+\sum_{r=1}^{q-1} A_{r,q}(1-\rho)^{r}\prod_{l=1}^{r}(s-1-l)\right]\,.
\label{eq:L}
\end{align}
For $s$-uniform hypergraphs, we give explicit formulas for $R(\rho)$ for a few specific parameter choices using Eq.~(\ref{eq:reduce_R}). These formulas are
 \begin{align}
	 R(\rho)&=\rho(1-\rho) \quad\textrm{for} \quad \textrm{either } q = 1 ~ \,\textrm{or}\, ~s = 2\,,
\nonumber \\
	R(\rho)&=\dfrac{\rho(1-\rho)}{(s-1)}\left[1+(s-2)\rho\right] \quad\textrm{for} \quad q = 2\,,
\nonumber \\
	R(\rho)&=\dfrac{\rho(1-\rho)}{(s-1)^2}\left[1+3(s-2)\rho+(s-2)(s-3)\rho^2 \right] \quad\textrm{for} \quad q = 3\,,
\nonumber \\
	R(\rho)&=\dfrac{\rho(1-\rho)}{(s-1)^3}\left[1+7(s-2)\rho+6(s-2)(s-3)\rho^2+(s-2)(s-3)(s-4)\rho^3 \right] \quad\textrm{for} \quad q = 4\,, \textrm{ and } 
\nonumber \\
	R(\rho)&=\dfrac{\rho(1-\rho)}{(s-1)^{q-1}}\left[1+\sum_{r=1}^{s-2} A_{r,q}\rho^r \prod_{l=1}^{r} (s-1-l)\right] \quad\textrm{for} \quad q \to \infty\,.
	\label{eq:v_rho_s5}
 \end{align}
When $s = N$ and $N\to\infty$, the raising operator $R(\rho)$ in Eq.~(\ref{eq:v_rho_s5}) becomes
 \begin{align}
	 R(\rho) = (1 - \rho)\rho^{q}\,.
	 \label{eq:v_rho_sN}
 \end{align}
In Fig.~\ref{eq:v_s=N}(a), we show the drift function $v(\rho) \equiv R(\rho) - L(\rho) = R(\rho) - R(1-\rho)$ for $s = 7$ from Eq.~(\ref{eq:v_rho_s5}). In Fig.~\ref{eq:v_s=N}(b), we show the drift function
 for $s = N \rightarrow \infty$.
When $s = N \rightarrow \infty$, the drift function is
\begin{align}\label{drifty}
	v(\rho) = (1 - \rho)\rho^q - \rho(1 - \rho)^q\,,
\end{align}
which corresponds to Eq.~(10) of Ref.~\cite{ramirez}. 
From Eq.~(\ref{eq:v_rho_s5}), we see that when either $q = 1$ (i.e., no nonlinearity) or $s = 2$ (i.e., no group interactions), $R(\rho) = L(\rho) = \rho(1 - \rho)$, which implies that the drift function $v(\rho) = 0$. The dynamics becomes purely diffusive, as in Ref.~\cite{sood-redner}. We are interested in the competition between group interactions and nonlinearity, so we focus our analysis on situations with $q \geq 2$ and $s \geq 3$ unless we note otherwise.

\subsection{Derivation of the exit probability $\Phi(\rho)$}
The exit probability $\Phi(\rho)$ satisfies the recursion relation 
\begin{align}
	\Phi(\rho) = R(\rho)\Phi(\rho+\delta\rho) + L(\rho)\Phi(\rho-\delta\rho)+[1-R(\rho)-L(\rho)]\Phi(\rho)\,,
\label{eq:exit_prob}
\end{align}
which is Eq.~(3) of the main manuscript. We Taylor-expand $\Phi(\rho\pm\delta\rho)$ in $\delta\rho$ up to second order and write
\begin{align}
	\Phi(\rho\pm\delta\rho) \approx \Phi(\rho)\pm\dfrac{\partial{\Phi(\rho)}}{\partial{\rho}}\delta\rho+\dfrac{1}{2}\dfrac{\partial^2{\Phi(\rho)}}{\partial{\rho^2}}(\delta\rho)^2\,.
\label{eq:talor2}
\end{align}
We then substitute Eq.~(\ref{eq:talor2}) into
Eq.~(\ref{eq:exit_prob}) to obtain the backward Kolmogorov equation 
\begin{align}
	v(\rho)\dfrac{\partial{\Phi(\rho)}}{\partial{\rho}} + D(\rho)\dfrac{\partial^2{\Phi(\rho)}}{\partial{\rho^2}}=0\,,
\label{eq:kolmo2}
\end{align}
where $v(\rho) = R(\rho) - L(\rho)$ is the drift function (which we defined in Sec.~\ref{transition})
and $D(\rho) \equiv [R(\rho) + L(\rho)]/(2N)$ is the diffusion function.
From Eqs.~(\ref{eq:reduce_R}) and (\ref{eq:L}), the coefficient of the drift term is
\begin{align}
	v(\rho) &= \dfrac{\rho(1-\rho)}{\sum_s sP(s)}\sum_s \dfrac{sP(s)}{(s-1)^{q-1}} \sum_{r=1}^{q-1} A_{r,q}\left[\rho^{r}-(1-\rho)^r\right]\prod_{l=1}^{r}(s-1-l)   \label{eq:rate_equation}
\end{align}
and the coefficient of the diffusion term is
\begin{align}
	D(\rho) &= \dfrac{1}{2N}\dfrac{\rho(1-\rho)}{\sum_s sP(s)}\sum_s \dfrac{sP(s)}{(s-1)^{q-1}} \left\{2+\sum_{r=1}^{q-1} A_{r,q}\left[\rho^r+(1-\rho)^{r}\right]\prod_{l=1}^{r}(s-1-l)\right\} \,. \label{eq:D_rho}
\end{align}

We now obtain an explicit expression for $\Phi(\rho)$ for the GVM with $q = 2$ on an annealed $s$-uniform hypergraph.
In this case, the raising and lowering operators are 
\begin{align}
	R(\rho) &= \dfrac{\rho(1-\rho)}{(s-1)}\left[1+(s-2)\rho\right] \,, \notag \\
	L(\rho) &= \dfrac{\rho(1-\rho)}{(s-1)}\left[1+(s-2)(1-\rho)\right]\,,
\end{align}
which implies that
\begin{align}
	v(\rho) &= \frac{(s-2)}{(s-1)}\rho(1-\rho)(2\rho-1) \,, \notag \\
	 D(\rho) &= \frac{1}{2N}\frac{s}{(s-1)}\rho(1-\rho) \,.
\end{align}
We write the derivative of $\Phi(\rho)$ with respect to $\rho$ as $\mathrm{exp}[-f(\rho)]$, where
\begin{align}
	f(\rho) &= \int \frac{v(\rho)}{D(\rho)}  d\rho \notag \\
	&=\int \dfrac{2N(s-2)(2\rho-1)}{s}  d\rho
\nonumber \\
	&=\dfrac{N(s-2)}{2s}-\dfrac{2N(s-2)(\rho-1/2)^2}{s} \,.
\end{align}
Using the boundary conditions $\Phi(0) = 0$ and $\Phi(1) = 1$, we obtain
\begin{align}
	\Phi(\rho)=\dfrac{\displaystyle\int_{0}^{\rho} \mathrm{exp}\left[\frac{-2N(s-2)(\rho'-1/2)^2}{s}\right] d\rho'}{\displaystyle\int_{0}^{1} \mathrm{exp}\left[\frac{-2N(s-2)(\rho'-1/2)^2}{s}\right] d\rho'} =\dfrac{\displaystyle\int_{-\frac{N(s-2)}{2s}}^{\frac{N(s-2)(\rho-1/2)}{2s}} \mathrm{exp}\left[-y^{2}\right] dy}{\displaystyle\int_{-\frac{N(s-2)}{2s}}^{\frac{N(s-2)}{2s}} \mathrm{exp}\left[-y^2\right] dy} \,,
\label{eq:exit_prob_final}
\end{align}
where $y = \sqrt{2N(s-2)/s}(\rho-1/2)$.
This yields
Eq.~(5) of the main manuscript:
\begin{align}
	\Phi(\rho)=\dfrac{1}{2}+\dfrac{\mathrm{erf}\left(\sqrt{\dfrac{2N(s-2)}{s}}\left(\rho-\dfrac{1}{2}\right)\right)}{2 \,\mathrm{erf}\left(\sqrt{\dfrac{N(s-2)}{2s}}\right)}\, ,
\label{eq:exit_prob_final2}
\end{align}
where $\mathrm{erf}(x) = 2\int_{0}^{x} \mathrm{exp}[-z^2] dz /\sqrt{\pi}$ is the error function.


\subsection{Derivations of the exit times $T(\rho)$ and $\tau \equiv T(\rho = 1/2)$}
\label{exitexit}

The exit time $T(\rho)$ satisfies the recursion relation
\begin{align}
	T(\rho) = R(\rho)T(\rho+\delta\rho) + L(\rho)T(\rho-\delta\rho) + [1-R(\rho)-L(\rho)]T(\rho) + \delta t \,,
\label{eq:recur}
\end{align}
{with $\delta t = 1/N$}. 
We Taylor-expand $T(\rho\pm\delta\rho)$ in $\delta\rho$ up to second order and write
\begin{align}
	T(\rho\pm\delta\rho) \approx T(\rho)\pm\dfrac{\partial{T(\rho)}}{\partial{\rho}}\delta\rho+\dfrac{1}{2}\dfrac{\partial^2{T(\rho)}}{\partial{\rho^2}}(\delta\rho)^2 \,, 
\label{eq:talor}
\end{align}
which we insert into Eq.~(\ref{eq:recur}) to obtain the backward Kolmogorov equation 
\begin{align}
	v(\rho)\dfrac{\partial{T(\rho)}}{\partial{\rho}}+D(\rho)\dfrac{\partial^2{T(\rho)}}{\partial{\rho^2}}=-1 \,,
\label{eq:kolmo}
\end{align}
which is Eq.~(6) of the main manuscript.


\subsubsection{Numerical solution of the recursion relation \eqref{eq:recur}}
\label{previous}

It is challenging to obtain an exact analytical solution of Eq.~(\ref{eq:kolmo}), so we compute
$T(\rho)$ by numerically solving Eq.~(\ref{eq:recur}). This numerical computation yields the plots in Figs.~2, 4 and 5 of the main manuscript.
We use discretized variables $X_m \equiv X(\rho = m/N)$, where $m \in \{0, 1, \ldots, N\}$.
Equation~(\ref{eq:recur}) then becomes 
\begin{align}
	-\dfrac{1}{N}=R_m Z_m-L_m Z_{m-1} \,,
\label{eq:reduced_recur}
\end{align}
where $Z_m \equiv T_{m+1} - T_m$. With the boundary conditions $T_m = T_{N - m}$ and $T_0 = T_N = 0$, we obtain
 $Z_{\frac{N}{2}-1} = \frac{1}{2NR_{\frac{N}{2}}}$ and $Z_{0} = T_1$.
We use Eq.~(\ref{eq:reduced_recur}) to determine 
$Z_m$ for the other values of $m$.
We obtain $T_m$ by calculating
\begin{align}
	T_m = \sum_{l=0}^{m-1} Z_l \,.
\end{align}


\subsubsection{Derivation of the logarithmic scalings of \,$T(\rho)$ and $\tau$}

The numerical computation of $T(\rho)$ in Sec.~\ref{previous} is useful, but it does not provide sufficient intuition about $T(\rho)$. To obtain such intuition, we perform an approximate analytical calculation by neglecting the second-order (i.e., diffusion) term in Eq.~(\ref{eq:kolmo}). The rationale behind neglecting the diffusion term is that the diffusion function $D(\rho)$, which includes the factor $1/N$, is much smaller than the drift function $v(\rho)$. 
We thus expect to extract the correct leading-order scaling for $T(\rho)$ under this approximation. 
With the boundary conditions $T(\rho = 0) = T(\rho = 1) = 0$, the solution of the approximation of
Eq.~(\ref{eq:kolmo}) satisfies
\begin{align}
	T(\rho) \approx \displaystyle\int_{\frac{1}{N}}^{\rho} \dfrac{-1}{v(\rho')} d\rho' \, ,
\label{eq:T_approx}
\end{align}
where we set the lower limit of the integral to $1/N$ to keep track of the $N$-dependence of $T(\rho)$. 
The drift function $v(\rho)$ is given by Eq.~(\ref{eq:rate_equation}). 
For an $s$-uniform hypergraph, Eq.~(\ref{eq:rate_equation}) is 
\begin{align}
	v(\rho)&=\dfrac{\rho(1-\rho)}{(s-1)^{q-1}} \sum_{r=1}^{q-1} A_{r,q}\left[\rho^{r}-(1-\rho)^r\right]\prod_{l=1}^{r}(s-1-l) \,.
\label{eq:rate_equation23}
\end{align} 
The factor $\rho^r-(1-\rho)^r$ in Eqs.~(\ref{eq:rate_equation}, \ref{eq:rate_equation23}) becomes $0$ only when $\rho = 1/2$ because
the function $\rho^r$ is a bijection. Furthermore, 
\begin{align*}
	\left.\frac{\rho^r-(1-\rho)^r}{(2\rho-1)}\right|_{\rho=\frac{1}{2}}=r\left(\frac{1}{2}\right)^{r-1}\,,
\end{align*}	
so $\rho = 1/2$ is a simple root. We thus write the drift function $v(\rho)$ as
\begin{align}
	v(\rho) = \rho(1 - \rho)(2\rho - 1)f(\rho,s,q) \,,
\label{eq:approx_rate}
\end{align}
where $f(\rho,s,q)$ does not contain real zeros of $\rho$. 
Performing a partial-fraction expansion of Eq.~(\ref{eq:approx_rate}) yields
\begin{align}
	\dfrac{1}{v(\rho)}=\dfrac{C_1(s,q)}{\rho}+\dfrac{C_2(s,q)}{(1-\rho)}+\dfrac{C_3(s,q)}{(2\rho-1)}+\dfrac{g(\rho,s,q)}{f(\rho,s,q)} \,, 
\label{eq:decom}
\end{align}
which we insert into Eq.~(\ref{eq:T_approx}) to obtain the approximate exit time. 
We first compute the exit time $\tau \equiv T(\rho = 1/2)$ for
the balanced initial condition (which has
the same number of nodes in each state). 
We obtain 
\begin{align}
	\tau(s,q,N) = T(\rho = 1/2) \approx \int_{\frac{1}{2}-\frac{1}{\sqrt{N}}}^{\frac{1}{N}} \frac{1}{v(\rho')} d\rho' \sim \left(-C_1(s,q) + \frac{C_3(s,q)}{4}\right)\ln N \equiv {\cal A}(s,q)\ln N \,, 
\label{eq:inte}
\end{align}
where we offset the initial density by $1/\sqrt{N}$ from $1/2$ both to avoid getting trapped at
the equilibrium point $\rho = 1/2$ and to account for 
stochasticity. The notation $\sim$ signifies leading-order scaling in $\ln N$.
The expression for $C_1(s,q)$ is 
\begin{align}
	C_1(s,q) \equiv \left.\dfrac{\rho}{v(\rho)}\right|_{\rho=0}=\dfrac{(s-1)^{q-1}}{-\sum\limits_{r=1}^{q-1} A_{r,q}\prod_{l=1}^{r}(s-1-l)}=\dfrac{(s-1)^{q-1}}{1-(s-1)^{q-1}} \,,
\end{align}
where the last equality follows from $\sum\limits_{r=1}^{q-1} A_{r,q}\prod_{l=1}^{r}(s-1-l)=\left.\frac{(s-1)^{q-1}R(\rho)}{\rho(1-\rho)}\right|_{\rho=1}-1=(s-1)^{q-1} - 1$ by using Eq.~(\ref{eq:n_q}). 
The expression for $C_3(s,q)$ is 
\begin{align}
	C_3(s,q) \equiv \left.\dfrac{(2\rho-1)}{v(\rho)}\right|_{\rho=\frac{1}{2}}=\dfrac{4(s-1)^{q-1}}{\sum\limits_{r=1}^{q-1} r\left(\frac{1}{2}\right)^{r-1} A_{r,q}\prod_{l=1}^{r}(s-1-l)} \,.
\end{align}
Therefore, the leading-order behavior of $\tau(s,q,N)$ as $N \rightarrow \infty$ is
\begin{align}
	\tau(s,q,N)\sim \left[\frac{(s-1)^{q-1}}{(s-1)^{q-1}-1}+\frac{(s-1)^{q-1}}{\sum\limits_{r=1}^{q-1} r\left(\frac{1}{2}\right)^{r-1} A_{r,q}\prod_{l=1}^{r}(s-1-l)}\right]\ln N \,,
\label{eq:leading}
\end{align}
which is one of the main theoretical results of our paper. It indicates
the logarithmic scaling of $\tau$ with $N$ for our
generic GVM for $q \ge 2$ and $s \ge 3$ 
on $s$-uniform hypergraphs. 


\subsubsection{Explicit derivation of the formulas for the exit time $\tau$ in the main manuscript}

We now obtain the explicit leading-order formulas for $\tau$ in the main manuscript from the general formula in Eq.~(\ref{eq:leading}). 
When $q = 2$, Eq.~(\ref{eq:leading}) becomes 
\begin{align}
	\tau(s,q=2,N) &\sim \left[\dfrac{(s-1)}{(s-2)}+\dfrac{(s-1)}{A_{1,2}(s-2)}\right]\ln N
\nonumber \\
	&=\dfrac{2(s-1)}{(s-2)}\ln N \,,
\label{eq:leading_q2}
\end{align}
which is Eq.~(8) of the main manuscript.
When $q = 5$, Eq.~(\ref{eq:leading}) becomes 
\begin{align}
	\tau(s,q=5,N)&\sim \left[\dfrac{(s-1)^{4}}{(s-1)^{4}-1}+\dfrac{(s-1)^{4}}{\sum\limits_{r=1}^{4} r\left(\frac{1}{2}\right)^{r-1} A_{r,5}\prod_{l=1}^{r}(s-1-l)}\right]\ln N \nonumber \\
&=\left\{\dfrac{(s-1)^{4}}{(s-1)^{4}-1}+\dfrac{(s-1)^{4}}{(s-2)\left[ A_{1,5}+(s-3)A_{2,5}+\frac{3}{4}(s-3)(s-4)A_{3,5}+\frac{1}{2}(s-3)(s-4)(s-5)A_{4,5}\right]}\right\}\ln N
\nonumber \\
	&=\dfrac{(s-1)^4(3s-4)(s+1)}{s(s-2)(s^2+3s-8)(s^2-2s+2)}\ln N \,,
\label{eq:leading_q5}
\end{align}
which is Eq.~(9) of the main manuscript. The exit time $\tau$ diverges for $s = 2$ in Eq.~(\ref{eq:leading}), so it also diverges in Eqs.~(\ref{eq:leading_q2}) and (\ref{eq:leading_q5}).
For $s = N$ with $N \gg 1$, the denominator of $C_3$ is dominated by the order-($q - 1$) term. That is, 
\begin{align*}
	\sum\limits_{r=1}^{q-1} r\left(\frac{1}{2}\right)^{r-1} A_{r,q}\prod_{l=1}^{r}(s-1-l) \approx \frac{(q-1)}{2^{q-2}}(s-1)^{q-1}\,. 
\end{align*}	
Therefore, for $s = N$, Eq.~(\ref{eq:leading}) becomes  
\begin{align}
	\tau(q,N)\sim \left[1+\dfrac{2^{q-2}}{(q-1)}\right]\ln N \,,
\label{eq:sN}
\end{align}
which is Eq.~(10) of the main manuscript. Equation~(\ref{eq:sN}) also applies to the $q$-voter model on complete (i.e., fully-connected) dyadic networks, and it agrees with the results for $q = 2$ and $q = 3$ 
in Ref.~\cite{ramirez}.
When $s$ is finite, it is convenient to replace the upper limit $r = q-1$ of the sum $\sum\limits_{r=1}^{q-1} r\left(\frac{1}{2}\right)^{r-1} A_{r,q}\prod_{l=1}^{r}(s-1-l)$ by $r = s - 2$. For example, when we do this, Eq.~(\ref{eq:leading}) becomes 
\begin{align}
	\tau(s=3,q,N)&\sim \left(\dfrac{2^{q-1}}{2^{q-1}-1}+\dfrac{2^{q-1}}{A_{1,q}}\right)\ln N
\nonumber \\
	&=\left(\dfrac{2^{q}}{2^{q-1}-1}\right)\ln N 
\end{align}	
for $s = 3$ and
\begin{align}
	\tau(s=5,q,N)&\sim \left[\dfrac{4^{q-1}}{4^{q-1}-1}+\dfrac{4^{q-1}}{\sum_{r=1}^{3} r\left(\frac{1}{2}\right)^{r-1} A_{r,q}\prod_{l=1}^{r} (4-l)}\right]\ln N 
\nonumber \\
	&=\left[\dfrac{4^{q-1}}{4^{q-1}-1}+\dfrac{4^{q}}{3(4^{q-1}+3^{q-1}-2^{q-1}-1)}\right]\ln N 
\nonumber \\
	&=\dfrac{4^{q-1}}{4^{q-1}-1}\left[\dfrac{\frac{7}{3}(4^{q-1}-1)+3^{q-1}-2^{q-1}}{4^{q-1}-1+3^{q-1}-2^{q-1}}\right]\ln N 
\end{align}
for $s = 5$.


\subsubsection{Dependence of the exit time $T(\rho_0)$ on the initial density $\rho_0$}

\begin{figure}[t]
\includegraphics[width=0.5\linewidth]{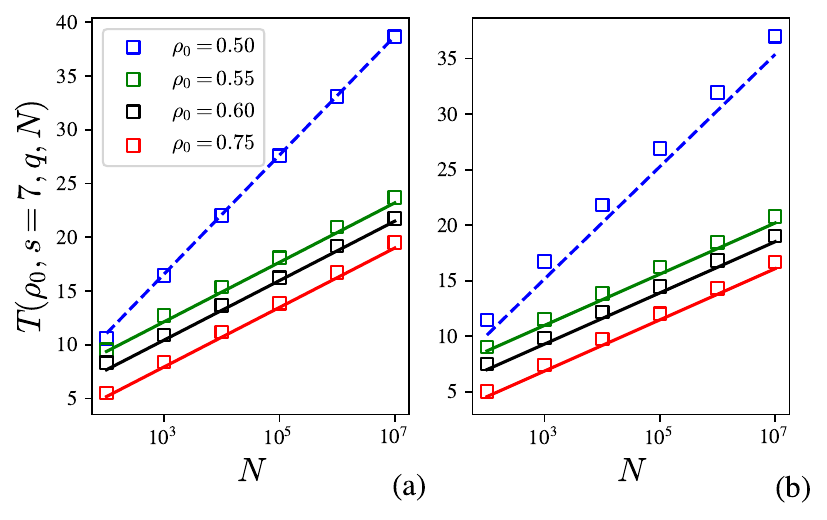}
\caption{The exit time $T(\rho_0)$ of the GVM on annealed $7$-uniform hypergraphs for different initial densities $\rho_0$ when (a) $q = 2$ and (b) $q = 5$.  The symbols give the means of $10^6$ (when $N \le 10^4$) or $10^3$ (when $N \ge 10^5$) independent MC simulations of the GVM on $N$-node hypergraphs.
The solid lines are theoretical results from (a) Eq.~(\ref{eq:T_q2_3}) and (b) Eq.~(\ref{eq:q_4_5}). We obtain the dashed lines from (a) Eq.~(\ref{eq:leading_q2}) and (b) Eq.~(\ref{eq:leading_q5}). 
}
\label{fig:initial-rho}
\end{figure}

We now compute the initial density-dependent exit time $T(\rho_0)$ under
the approximation of Eq.~(\ref{eq:T_approx}). From Eqs.~(\ref{eq:T_approx}) and (\ref{eq:decom}), 
the leading-order expression for $T(\rho_0)$ for the initial density $\rho_0$
away from $\rho_0 = 1/2$ takes the form $T(\rho_0)\sim  -C_1\ln N + T_0(\rho_0)$. 
That is, it scales as
$\ln N$
with a $\rho_0$-independent 
amplitude $-C_1$ and $\rho_0$-dependent integration constant $T_0(\rho_0)$. 
We elaborate on the derivation of $T(\rho_0)$ for several
values of $q$.
For $q = 2$ and $q = 3$, the drift function is
\begin{align}
	v(\rho) = \dfrac{(s-1)^{q-1}-1}{(s-1)^{q-1}}\rho(1-\rho)(2\rho-1) \,,
\label{eq:eq_q2_3}
\end{align}
from which we obtain 
\begin{align}
	T(\rho_0,s,q)& \approx \int_{\rho_0}^{\frac{1}{N}} \dfrac{(s-1)^{q-1}}{\left[(s-1)^{q-1}-1\right]\rho'(1-\rho')(2\rho'-1)} d\rho' \nonumber\\
	&=\dfrac{(s-1)^{q-1}}{\left[(s-1)^{q-1}-1\right]}\ln\left[\dfrac{(1-\frac{2}{N})^2\rho_0(1-\rho_0)}{(1-2\rho_0)^2\frac{1}{N}(1-\frac{1}{N})}\right] \nonumber \\
	&\sim \dfrac{(s-1)^{q-1}}{\left[(s-1)^{q-1}-1\right]}\ln\left[\dfrac{N\rho_0(1-\rho_0)}{(1-2\rho_0)^2}\right]\,, 
\label{eq:T_q2_3}
\end{align}
where the last step uses the fact that $N \gg 1$. 
Equation~(\ref{eq:T_q2_3}) with $s \to \infty$ is equivalent to Eq.~(17) of Ref.~\cite{ramirez}.
For $q = 4$ and $q = 5$, the drift function is
\begin{align}
	v(\rho) = \rho(1-\rho)(2\rho-1)\left[a(s,q)\rho^2-a(s,q)\rho+b(s,q)\right] \,,
\label{eq:eq_q4_5}
\end{align}
where $a(s,q=4)=\frac{(s-2)(s-3)(s-4)}{(s-1)^3}$, $a(s,q=5)=\frac{2s(s-2)(s-3)(s-4)}{(s-1)^4}$, and $b(s,q)=\frac{(s-1)^{q-1}-1}{(s-1)^{q-1}}$. The exit time
$T(\rho_0>1/2)$ is 
\begin{align}
	T(\rho_0,s,q) &\approx \int_{1-\rho_0}^{\frac{1}{N}} \dfrac{d\rho'}{\rho'(1-\rho')(2\rho'-1)\left[a(s,q)\rho'^2-a(s,q)\rho'+b(s,q)\right]} \nonumber
\\ \nonumber
&=\int_{1-\rho_0}^{\frac{1}{N}} \dfrac{1}{b(s,q)}\left[\dfrac{1}{(1-\rho')}-\dfrac{1}{\rho'}\right]+\dfrac{4}{\left[b(s,q)-\frac{a(s,q)}{4}\right](2\rho'-1)}+\left[\dfrac{1}{b(s,q)[1-\frac{4b(s,q)}{a(s,q)}]}\right]\dfrac{2\rho'-1}{\rho'^2-\rho'+\frac{b(s,q)}{a(s,q)}} d\rho' \\ \nonumber
	&=\dfrac{1}{b(s,q)}\ln\left[\dfrac{(1-\rho_0)\rho_0}{\left(1-\frac{1}{N}\right)\frac{1}{N}}\right]+\dfrac{2}{\left[b(s,q)-\frac{a(s,q)}{4}\right]}\ln\left(\frac{1-\frac{2}{N}}{2\rho_0-1}\right)
+\dfrac{1}{b(s,q)[1-\frac{4b(s,q)}{a(s,q)}]}\ln\left[\dfrac{\frac{1}{N^2}-\frac{1}{N}
+\frac{b(s,q)}{a(s,q)}}{\rho_0^2-\rho_0+\frac{b(s,q)}{a(s,q)}}\right] \\ 
	&\sim  \dfrac{1}{b(s,q)}\ln[N\rho_0(1-\rho_0)]-\dfrac{2}{\left[b(s,q)-\frac{a(s,q)}{4}\right]}\ln(2\rho_0-1)+\dfrac{1}{b(s,q)[1-\frac{4b(s,q)}{a(s,q)}]}\ln\left[\dfrac{
	\frac{b(s,q)}{a(s,q)}}{\rho_0^2-\rho_0+\frac{b(s,q)}{a(s,q)}}\right] 
\label{eq:q_4_5}
\end{align}
when $N \gg 1$. We confirm Eqs.~(\ref{eq:T_q2_3}) and (\ref{eq:q_4_5}) using MC simulations (see Fig.~\ref{fig:initial-rho}).

{\section{Special cases and variants of our GVM}}
In this section, we discuss several special cases and variants of our GVM.

\medskip
\medskip

{
\subsection{GVM with nonlinearity strength $q = 1$}

\vspace{-2 mm}

On an $s$-uniform hypergraph, the GVM with $q = 1$ is equivalent to a VM on a dyadic network for all hyperedge sizes $s$.
We now verify this statement. From Eq.~(\ref{eq:v_rho_s5}), when
$q = 1$, we have $R(\rho) = L(\rho) = \rho(1 - \rho)$ for all $s$. That is, the raising and lowering operators
are the same as those for a
VM on a dyadic network \cite{sood-redner}. 
Therefore, we obtain the same backward Kolmogorov equation,
\begin{align} \label{eq:reduced_kolmo}
	\dfrac{\rho(1 - \rho)}{N}\dfrac{\partial^2{T(\rho)}}{\partial{\rho^2}} = -1 \,,
\end{align}
for $T(\rho)$. We thus also obtain the same solution
\begin{align}
	T(\rho) = N\left[\rho\ln{\dfrac{1}{\rho}} + (1 - \rho)\ln\left(\dfrac{1}{1 - \rho}\right)\right] 
	\propto N \,.
\end{align}
We thereby obtain the exit time $\tau = T(\rho = 1/2) = N \ln 2$, as in Ref.~\cite{sood-redner}.


{\subsection{{GVM with geometric $P(s)$}}
We examine the GVM on annealed hypergraphs with a geometric hyperedge-size distribution $P(s)$. The formula for the distribution is $P(s) = \frac{1}{\langle s\rangle - 1}\left(\frac{\langle s\rangle - 2}{\langle s\rangle - 1}\right)^{s - 2}$, where $\langle s\rangle$ is the mean hyperedge size, which we also use
for the simplicial GVM and Fig.~2(a) in the main manuscript. In Fig.~\ref{fig:geometric}, we show the exit time $\tau$ from MC
simulations and the recursion relation (\ref{eq:recur}) 
for several values of the group size} $\langle s \rangle$. As in Fig.~5(a) of the main manuscript, the exit time $\tau$ exhibits optimality. 

\begin{figure}[h]
\includegraphics[width=.5\linewidth]{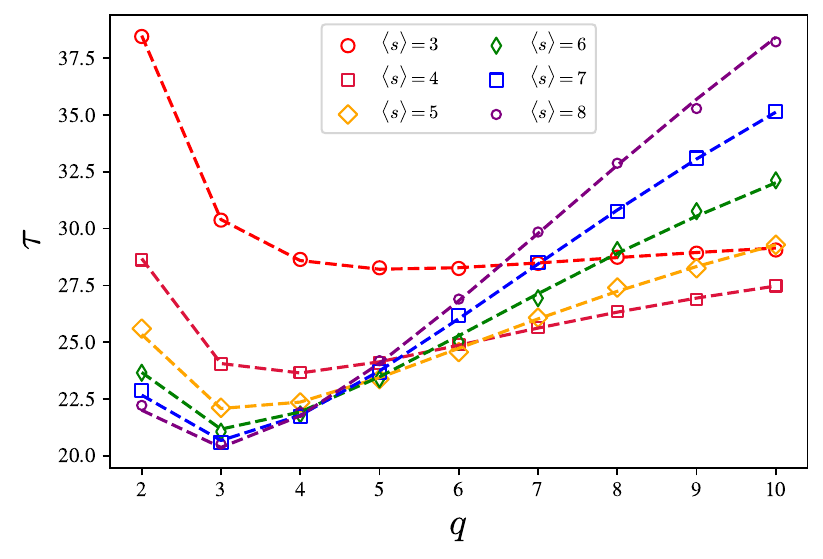}
\caption{{The dependence on the nonlinearity strength $q$ of the exit time $\tau$ for the GVM on hypergraphs with $N = 10^4$ nodes with a geometric $P(s)$ for several values of the mean group size $\langle s \rangle$. The markers are means of $10^4$ independent MC simulations of the GVM on annealed hypergraphs, and the analytical curves are from numerical solutions of the recursion relation (\ref{eq:recur}). 
}}
\label{fig:geometric}
\end{figure}

}

\newpage
\subsection{GVM with power law $P(k)$}
\begin{figure}[t]
\includegraphics[width=.5\linewidth]{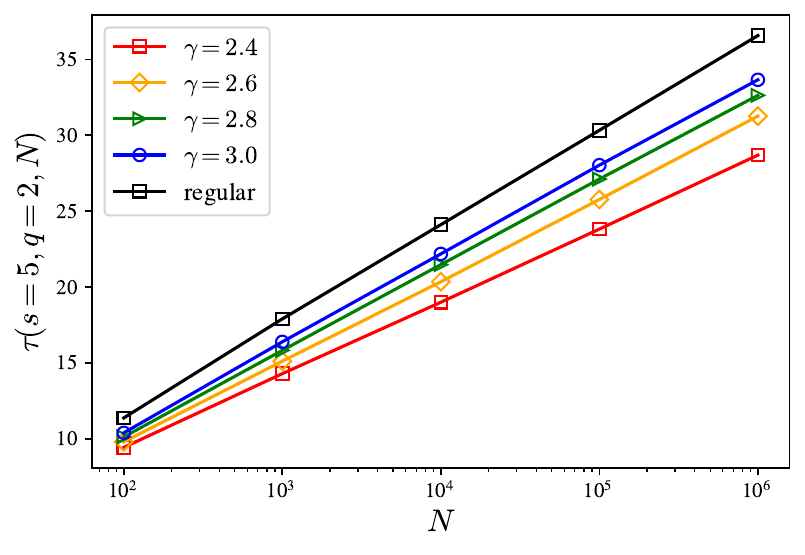}
\caption{The dependence on the hypergraph size $N$ of the exit time $\tau$ for the GVM with power law $P(k)$ for several values of the power-law exponent $\gamma$. The markers are means of $10^3$ independent MC simulations of the GVM with nonlinearity strength $q = 2$ on annealed $5$-uniform hypergraphs. We draw curves between these points as visual guides.
The ``regular" case corresponds to $\gamma \to \infty$ and reduces to the GVM that we studied in the main manuscript.
We use semilogarithmic coordinates, and we observe that
the exit time $\tau$ scales logarithmically in $N$.
}
\label{fig:pk}
\end{figure}

We examine the GVM with a power-law degree distribution $P(k)$ on annealed $s$-uniform hypergraphs by performing MC simulations. (The degree $k$ of a node is the number of hyperedges it is in.) The degree distribution is $P(k) = k^{-\gamma}/\sum_{\ell=1}^{N}\ell^{-\gamma}$, where $k \in \{1,2, \ldots,N\}$.
The degree of each node is a quenched (i.e., fixed) random variable that follows the distribution $P(k)$. 
To account for the degree distribution $P(k)$, we modify stage (ii) of the MC step in Sec.~\ref{stages}
to the following:
\begin{itemize}
\item[(ii)] We select $s - 1$ distinct nodes, each with a probability that is proportional to its degree, to form a hyperedge $h$.
\end{itemize}
In Fig.~\ref{fig:pk}, we show the results of our MC simulations.
In these simulations, we take $s = 5$ and $q = 2$.
As in Fig.~4 of the main manuscript, the exit time $\tau$ scales with $\ln N$. However, the prefactor of $\ln N$ decreases as $\gamma \downarrow 2$. In a VM on a dyadic network, degree heterogeneity affects the exit time $\tau$ significantly~\cite{sood-redner}. Accordingly, in future work, it is worthwhile to analytically investigate the impact of degree heterogeneities on exit time.


\subsection{Simplicial GVM}
In the simplicial GVM, a node flips its state
in stage (iii) if the states
of all of its $s - 1$ neighbors are different from its state.
Therefore, the raising and the lowering operators are
\begin{align}
	R(\rho) &= \frac{\sum_s sP(s)(1-\rho)\rho^{s-1}}{\sum_s sP(s)}\,, \notag  \\
	L(\rho) &= \frac{\sum_s sP(s)\rho(1-\rho)^{s-1}}{\sum_s sP(s)} \,.
\end{align}
One can insert the transition probabilities $R(\rho)$ and $L(\rho)$ into Eq.~(\ref{eq:recur}) to obtain a recursion relation for the exit time $T(\rho)$. 
To obtain a leading-order approximation of the exit time $\tau$, we write the drift function $v(\rho)$ as 
\begin{align}
	v(\rho) = \dfrac{\rho(1-\rho)}{\sum_s sP(s)}\sum_s sP(s)\left[\rho^{s-2} - (1-\rho)^{s-2}\right] \,,
\label{eq:simplicial_rate}
\end{align}
from which we obtain 
\begin{align}
	\tau &\sim \left[-\left.\dfrac{\rho}{v(\rho)}\right|_{\rho=0}+\left.\dfrac{2\rho-1}{4v(\rho)}\right|_{\rho=\frac{1}{2}}\right]\ln N  \nonumber \\
	&=\left[\dfrac{1}{1-\dfrac{2P(2)}{\langle s \rangle}}+\dfrac{\langle s \rangle}{{\sum\limits_{s=3}^N} s(s-2)\left(\dfrac{1}{2}\right)^{s-3}P(s)}\right]\ln N \,.
\label{eq:simplicial_tau}
\end{align}
As we can see in Eq.~(\ref{eq:simplicial_tau}), the exit time again scales logarithmically with the system size $N$.
For an $s$-uniform hypergraph with $s \ge 3$, we obtain
\begin{align} 
	\tau \sim \left[1+\frac{2^{s-3}}{(s-2)}\right]\ln N \,.
	\label{eq:tau_sim_suni}
\end{align} 
Equation~(\ref{eq:sN}) with $q = s - 1$ reduces to Eq.~(\ref{eq:tau_sim_suni}).
That is, for $N \to \infty$, the exit time $\tau$ is the same for the simplicial GVM on annealed $s$-uniform hypergraphs and the GVM with duplicate selection allowed and $q = s - 1$ on $N$-uniform hypergraphs.
In Fig.~2 of the main manuscript, we showed the results of MC simulations of the simplicial GVM on annealed hypergraphs with two different hyperedge-size distributions $P(s)$. We now compare these simulation
results with analytical results. 
First, we consider the geometric hyperedge-size distribution $P(s) = \frac{1}{\langle s\rangle-1}\left(\frac{\langle s\rangle-2}{\langle s\rangle-1}\right)^{s-2}$ for
$s \ge 2$ with mean hyperedge size $\langle s\rangle$. For this distribution, Eq.~(\ref{eq:simplicial_tau}) becomes
\begin{align} 
	\tau \sim \left(\frac{\langle s\rangle}{\langle s\rangle-2}\right)\left[\frac{\langle s\rangle-1}{\langle s\rangle+1}+\frac{\langle s\rangle^3}{4(5\langle s\rangle-4)}\right]\ln N \,,
\label{eq:tau_geometric}
\end{align} 
which predicts that the exit time $\tau$ exhibits optimality with a minimum value at $\langle s\rangle^*\approx 3.58$. 
Second, we consider the power-law hyperedge-size distribution with exponent $\alpha$. This distribution has the formula 
$P(s) = \frac{s^{-\alpha}}{\zeta(\alpha)-1}$ for $s \ge 2$ and 
mean hyperedge size $\frac{\zeta(\alpha - 1) - 1}{\zeta(\alpha)-1}$, where $\zeta(z)$ is the Riemann zeta function. For this distribution, Eq.~(\ref{eq:simplicial_tau}) becomes 
\begin{align} 
	\tau &\sim \left(\sum_{s=2}^Ns^{1-\alpha}\right)\left[\frac{1}{\sum_{s=3}^Ns^{1-\alpha}}+\frac{1}{\sum_{s=3}^{N}\frac{(s-2)s^{1-\alpha}}{2^{s-3}}}\right] \ln N \nonumber \\
	&\approx\left[\zeta(\alpha-1)-1\right]\left[\frac{1}{\zeta(\alpha-1)-1-2^{1-\alpha}}+\frac{1}{8 \, \textrm{Li}_{\alpha-2}\left(\frac{1}{2}\right)-16 \, \textrm{Li}_{\alpha-1}\left(\frac{1}{2}\right)+4}\right]\ln N \, ,
\label{eq:tau_powerlaw}
\end{align} 
where $\textrm{Li}_{s}(z)$ is the polylogarithm function and the last step follows by taking the upper limit of the sums to $\infty$. 
Equation~\eqref{eq:tau_powerlaw} predicts that the exit time $\tau$ exhibits optimality with a minimum value at $\alpha^* \approx 2.87$.
As we showed in Fig.~2 of the main manuscript, our theoretical approximation is in good agreement with the results of MC simulations. The numerical solution of the recursion relation \eqref{eq:recur} (solid curves) agrees very well with the results of MC simulations, and the leading-order approximations (\ref{eq:tau_geometric}, \ref{eq:tau_powerlaw})
(dotted curves) successfully account for the 
optimality in $\tau$. 
The expression Eq.~(\ref{eq:simplicial_tau}) for $\tau$ for the simplicial GVM diverges when 
$\langle s\rangle\downarrow 2$ 
because the network reduces to a
dyadic network and $\tau$ crosses over to the diffusive behavior ${\cal O}(N)$ in this limit. 
The divergence of the exit time $\tau$ occurs in the $\langle s\rangle \downarrow 2$ limit of Eq.~(\ref{eq:tau_geometric}) and {in} the $\alpha \to \infty$ limit of Eq.~(\ref{eq:tau_powerlaw}).


\subsection{GVM without duplicate selections} \label{sec-s3b}
\begin{figure}[t]
\includegraphics[width=.5\linewidth]{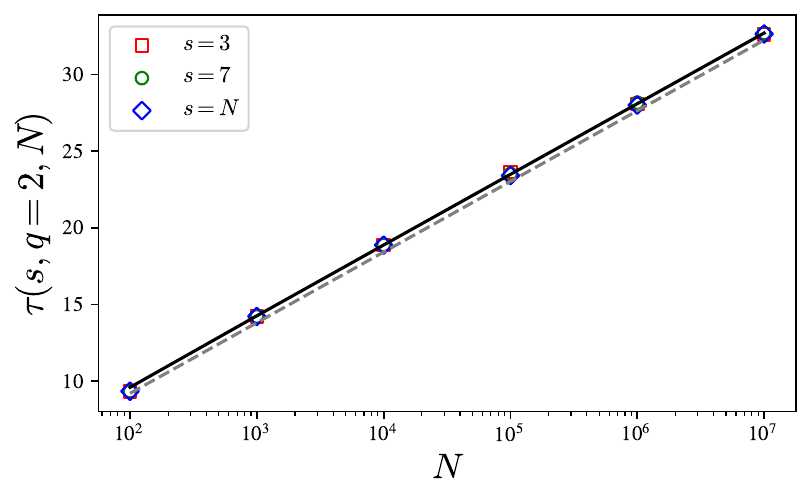}
\caption{Dependence of the exit time $\tau$ on the system size $N$ for a GVM without duplicate selections for nonlinearity strength $q=2$ and hyperedge sizes $s = 3$, $s = 7$, and $s = N$.
The markers indicate the means of $10^6$ (when $N \le 10^4$) or $10^3$ (when $N \ge 10^5$) independent MC simulations of this GVM on annealed $s$-uniform hypergraphs. The lines are solutions of the recursion relation \eqref{eq:recur} (solid) and the leading-order solution~(\ref{eq:tau_nodup}) (dotted). 
}
\label{fig:gvm_wo_dup}
\end{figure}

Our analysis also applies if we disallow duplicate selections in the neighbor-selection stage (iii) of the GVM (see Sec.~\ref{stages}).
In this case, the raising operator $R(\rho)$ and lowering operator $L(\rho)$ are 
\begin{align}
\label{eq:R_dup}
	R(\rho)&=\dfrac{(1-\rho)}{\sum_s sP(s)}\sum_s sP(s)\sum_{n={q}}^{s-1} \dfrac{(s-1)!}{n!(s-1-n)!}\rho^n (1-\rho)^{s-1-n}\dfrac{\dfrac{n!}{q!(n-q)!}}{\phantom{a}\dfrac{(s-1)!}{q!(s-1-q)!}\phantom{a}}
\nonumber \\
	&=\dfrac{(1-\rho)}{\sum_s sP(s)}\sum_s sP(s)\sum_{n={q}}^{s-1} \dfrac{(s-1-q)!}{(n-q)!(s-1-n)!}\rho^n (1-\rho)^{s-1-n}
\nonumber \\
	&=\dfrac{(1-\rho)}{\sum_s sP(s)}\sum_{s=q+1}^{\infty} sP(s)\rho^q \,,
\\
	L(\rho)&=\dfrac{\rho}{\sum_s sP(s)}\sum_{s=q+1}^{\infty} sP(s)(1-\rho)^q \,.
\end{align}
For $s$-uniform hypergraphs, Eq.~(\ref{eq:R_dup}) becomes 
\begin{align} 
	R(\rho) = \left\{\begin{array}{ll} (1-\rho)\rho^q \,, & \quad s\ge q+1 \\ 0 \,, & \quad s\le q \,. \end{array}\right. 
\end{align}
The model is meaningful only for $s \ge q+1$. In this case, $R(\rho)$ is the same as in Eq.~(\ref{eq:v_rho_sN}), which is for the GVM \textit{with duplicate selections allowed} on annealed $N$-uniform hypergraphs. 
  Therefore, we obtain the same exit time $\tau$ as in Eq.~(\ref{eq:sN}). This exit time is
\begin{align}
	\tau(q,N)\sim\left[1+\dfrac{2^{q-2}}{(q-1)}\right]\ln N \,.
\label{eq:tau_nodup}
\end{align}
In Fig.~\ref{fig:gvm_wo_dup}, we show results for $q = 2$, which has the exit time $\tau \sim 2 \ln N$.


\subsection{GVM with edge-update dynamics}
\begin{figure}[t]
\includegraphics[width=.55\linewidth]{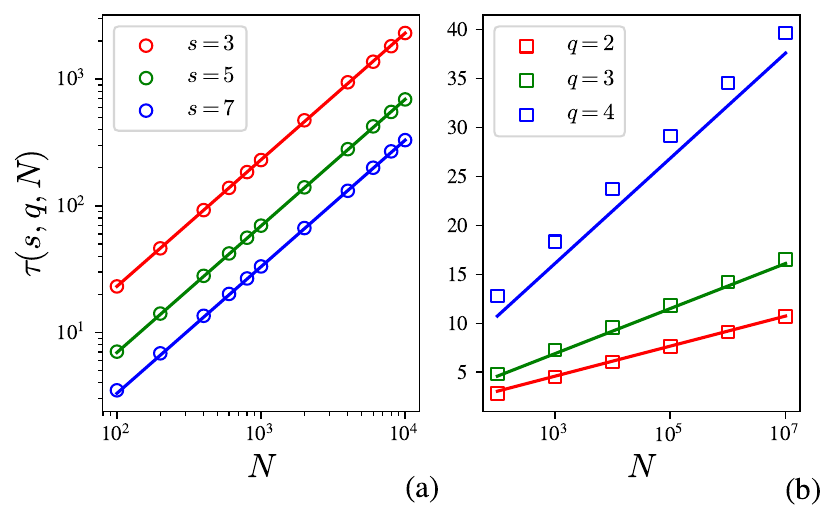}
\caption{The dependence of $\tau(s,q,N)$ on the hypergraph size $N$. The markers indicate means of $10^3$ independent
MC simulations of the GVM with edge-update dynamics on annealed $s$-uniform hypergraphs. 
In (a), we give results for the nonlinearity strength $q = 1$. The lines are solutions of $\tau(s,q=1,N)=\dfrac{2\ln2}{s(s-1)}N$ from Eq.~(\ref{eq:Nscale}) for different values of the hyperedge size $s$. In (b), we give results for nonlinearity strengths $q \geq 2$ and hyperedge size $s = 5$.
We obtain the lines from Eq.~(\ref{eq:log_scaling_hyperedge}).
}
\label{fig:hyperedge_fliping}
\end{figure}

One can frame standard VM dynamics in terms of edge-update rules, rather than the node-update rules that we discussed in the main manuscript. 
For edge-update dynamics on a dyadic network, in each step of an MC simulation, one (i) chooses an edge uniformly at random and (ii) updates the states of both of its attached edges to the same uniformly-randomly-chosen state when they have different states~\cite{suchecki}.
An equivalent way to implement (ii) is to select a uniformly random node that is attached to the edge and copy its state to the other node. 
We generalize edge-update dynamics to hypergraphs 
by considering the following GVM with edge-update dynamics. At each time step, an MC simulation has
the following stages: 
\begin{itemize}
\item[(i)] We select a size-$s$ hyperedge $h$ uniformly at random.  
\item[(ii-a)] We select $q$ distinct nodes uniformly at random from the $s$ nodes in the hyperedge $h$. 
\item[(ii-b)] If all of the $q$ nodes that we select in (ii-a) have the same 
state, we copy this state to every node in the hyperedge $h$. 
\end{itemize}
In stage (ii-b), the nodes with a state that is different from that of the $q$ nodes flip their state. In general, a time step can include more than one such node,
which is a crucial distinction from the node-update GVM 
in the main manuscript. Additionally, in (ii-a), we disallow duplicate selections of nodes. We make this choice
for technical convenience; one obtains qualitatively similar results if one allows duplicate selections.
From the model definition, we can readily write the transition probabilities for our edge-update GVM.
 The raising operator $R_{s,n}(\rho) \equiv P(\rho\rightarrow\rho+\delta\rho^{+}_{s,n})$, with $\delta\rho^{+}_{s,n}\equiv (s-n)/N$, is the transition probability that a hypergraph has $s - n$ more nodes in state $\mathbf{1}$ after a time step. The raising operator is given by
\begin{align}
	R_{s,n}(\rho) &= P(s)\dfrac{s!}{n!(s-n)!}\rho^{n} (1-\rho)^{s-n}\dfrac{\phantom{a}\dfrac{n!}{q!(n-q)!}\phantom{a}}{\dfrac{s!}{q!(s-q)!}} \nonumber\\
	&= P(s)\dfrac{(s-q)!}{(s-n)!(n-q)!}\rho^{n}(1-\rho)^{s-n} \,.
\label{eq:R_hyperedge}
\end{align}
The lowering operator $L_{s,n}(\rho) \equiv P(\rho\rightarrow\rho-\delta\rho^{-}_{s,n})$, with $\delta\rho^{-}_{s,n}\equiv n/N$, is the transition probability that a hypergraph has $n$ more nodes in state $\mathbf{0}$ after a time step. The lowering operator is given by
\begin{align}
	L_{s,n}(\rho)&= P(s)\dfrac{s!}{n!(s-n)!}\rho^{n} (1-\rho)^{s-n}\dfrac{\phantom{a}\dfrac{(s-n)!}{q!(s-n-q)!}\phantom{a}}{\dfrac{s!}{q!(s-q)!}} \nonumber\\
&=P(s)\dfrac{(s-q)!}{(s-n-q)!n!}\rho^{n}(1-\rho)^{s-n} \,.
\label{eq:L_hyperedge}
\end{align}
The recursion relation for the exit time $T(\rho)$ is 
\begin{align}
	T(\rho)=\sum_{s,n} \left[R_{s,n}(\rho)T(\rho+\delta\rho^{+}_{s,n})+L_{s,n}(\rho)T(\rho-\delta\rho^{-}_{s,n})\right]+\left[1-\sum_{s,n} R_{s,n}(\rho)+L_{s,n}(\rho)\right]T(\rho)+\delta t \,,
	\label{eq:recursion_hyperedge}
\end{align}
which yields the backward Kolmogorov equation
\begin{align}
	-1 &= \left[\sum_{s} P(s) (s-q)\{(1-\rho)\rho^q-\rho(1-\rho)^q\}\right]\dfrac{\partial{T(\rho)}}{\partial{\rho}}
\nonumber \\
	&\quad \,\,\,+ \left[\sum_{s} P(s)\dfrac{(s-q)}{2N}\{(s-q-1)\rho^q (1-\rho)^2+\rho^q (1-\rho)+(s-q-1)(1-\rho)^q \rho^2+(1-\rho)^q \rho\}\right]\dfrac{\partial^2{T(\rho)}}{\partial{\rho^2}} \nonumber
\\
	&\equiv \, v_h(\rho)\dfrac{\partial{T(\rho)}}{\partial{\rho}}+D_h(\rho)\dfrac{\partial^2{T(\rho)}}{\partial{\rho^2}}\,. 
\label{eq:kolmo_hyperedge}
\end{align}
In our derivation of Eq.~\eqref{eq:kolmo_hyperedge} from Eq.~\eqref{eq:recursion_hyperedge}, we use the Taylor expansion
\begin{align}
	T(\rho\pm\delta\rho^{\pm}_{s,n}) \approx T(\rho)\pm\dfrac{\partial{T(\rho)}}{\partial{\rho^{\pm}_{s,n}}}\delta\rho^{\pm}_{s,n}+\dfrac{1}{2}\dfrac{\partial^2{T(\rho)}}{\partial{\rho^2}}(\delta\rho^{\pm}_{s,n})^2 \,,
\label{eq:talor_hyperedge}
\end{align}
which we truncate after the second-order term.
When $q = 1$, the drift function $v_h(\rho) = 0$, so Eq.~(\ref{eq:kolmo_hyperedge}) reduces to
\begin{align}\label{this2}
	-1 = \dfrac{(\langle s^2 \rangle-\langle s \rangle)\rho(1-\rho)}{2N}\dfrac{\partial^2{T(\rho)}}{\partial{\rho^2}} \,,
\end{align}
where
$\langle s^r \rangle \equiv \sum_s s^r P(s)$. The solution of Eq.~\eqref{this2} is 
\begin{align}
	T(\rho) = \dfrac{2N}{(\langle s^2 \rangle-\langle s \rangle)}\left[\rho\ln{\dfrac{1}{\rho}}+(1-\rho)\ln\left(\dfrac{1}{1-\rho}\right)\right] \,.
\end{align}	
The exit time $\tau$ is 
\begin{align}
	\tau = T(\rho = 1/2)=\frac{2\ln 2}{\langle s^2\rangle - \langle s\rangle}N\propto N \,.
\label{eq:Nscale}
\end{align} 
In Fig.~\ref{fig:hyperedge_fliping}(a), we confirm Eq.~(\ref{eq:Nscale}). 
When $s = 2$, the exit time Eq.~(\ref{eq:Nscale}) reduces to the exit time for VM dynamics on a dyadic network.
When $q \geq 2$, Eq.~(\ref{eq:kolmo_hyperedge}) is 
not analytically solvable. Therefore, we apply the same approximation procedure as in Sec.~\ref{exitexit} and
obtain an approximate expression for
$\tau$ by substituting $v_h(\rho)$ 
into Eq.~(\ref{eq:inte}) and keeping the leading-order terms. 
The exit time $\tau$ is then
\begin{align}
	\tau(q \geq 2)\sim \dfrac{1}{\sum_{s \geq q + 1} P(s) (s - q)}\left[1+\dfrac{2^{q-2}}{(q-1)}\right]\ln N \,,
\label{eq:log_scaling_hyperedge}
\end{align}
which again scales logarithmically in $N$.
In Fig.~\ref{fig:hyperedge_fliping}(b), we compare this analytical prediction with the results of MC simulations on annealed $s$-uniform hypergraphs.
We obtain reasonable agreement.


